\documentclass[twocolumn,fleqn,usenatbib]{mnras}

\usepackage[T1]{fontenc}
\usepackage{ae,aecompl}
\usepackage{multirow}
\usepackage{booktabs}
\usepackage{tablefootnote}
\usepackage{booktabs,caption}

\usepackage[flushleft]{threeparttable}

\usepackage{graphicx}	
\usepackage{amsmath}	
\usepackage{amssymb}	
\usepackage[utf8]{inputenc}
\usepackage[dvipsnames]{xcolor}

\title[Kilonova luminosity function]{The luminosity functions of kilonovae from binary neutron star mergers under different equation of states}

\author[Zhao et al.]{
Chunyang Zhao,$^{1,2}$,
Youjun Lu,$^{1,2}$\thanks{luyj@nao.cas.cn}
Qingbo Chu$^{1,2}$,
and
Wen Zhao$^3$
\\
$^{1}$National Astronomical Observatories, Chinese Academy of Sciences, 20A Datun Road, Beijing, 100101, China;\\
$^{2}$School of Astronomy and Space Science, University of Chinese Academy of Sciences, No. 19A Yuquan Road, Beijing, 100049, China\\
$^3$Department of Astronomy, University of Science and Technology of China, Hefei, 230026, China
}

\begin{document}


\maketitle

\begin{abstract}
Kilonovae produced by mergers of binary neutron stars (BNSs) are important transient events to be detected by time domain surveys with the alerts from the ground-based gravitational wave detectors. The observational properties of these kilonovae depend on the physical processes involved in the merging processes and the equation of state (EOS) of neutron stars (NSs). In this paper, we investigate the dependence of kilonova luminosities on the parameters of BNS mergers, and estimate the distribution functions of kilonova peak luminosities (KLFs) at the u, g, r, i, y, and z bands as well as its dependence on the NS EOS, by adopting a comprehensive semi-analytical model for kilonovae (calibrated by the observations of GW170817), a population synthesis model for the cosmic BNSs, and the ejecta properties of BNS mergers predicted by numerical simulations. We find that the kilonova light curves depend on both the BNS properties and the NS EOS, and the KLFs at the considered bands are bimodal with the bright components mostly contributed by BNS mergers with total mass $\lesssim 3.2M_\odot$/$2.8M_\odot$ and fainter components mostly contributed by BNS mergers with total mass $\gtrsim 3.2M_\odot$/$2.8M_\odot$ by assuming a stiff/soft (DD2/SLy) EOS. The emission of the kilonovae in the KLF bright components is mostly due to the radiation from the wind ejecta by the remnant discs of BNS mergers, while the emission of the kilonovae in the KLF faint components is mostly due to the radiation from the dynamical ejecta by the BNS mergers.
\end{abstract}




\section{Introduction}
\label{sec:intro}

The first detection of gravitational waves (GWs) from a binary neutron star (BNS) merger (GW170817; \citealt{Abbott-2017-161101-161101}) and its multiwavelength electromagnetic (EM) counterparts \citep[e.g.,][]{Coughlin-2019-91-96} mark the beginning of a new era of multimessenger astronomy. It has already provided invaluable information to the physical processes related to BNS mergers, the nature of neutron stars (NSs), and the cosmological implications of these sources as standard sirens \citep[e.g.,][]{Abbott-2017-13-13, Abbott-2018-161101-161101, Abbott-2019-11001-11001, Kasen-2017-80-84, Margalit-2017-19-19, Chen2018545547, Mandel20181212,  Hotokezaka2019940944, Dietrich202014501453, Howlett202038033815, Nakar2021114114}. The laser interferometer gravitational wave observatory (LIGO) and Virgo detected the second BNS merger, GW190425, during its third operation (O3) period but did not detect its EM counterparts \citep{Abbott-2020-3-3}, partly because of its large distance and poor sky localization  \citep[][]{Coughlin-2019-19-19, Hosseinzadeh-2019-4-4}. These GW detected events have put a strong constraint on kilonova properties \citep{Abbott-2017-13-13} and the BNS merger rate $\sim 320{\rm Gpc}^{-3}{\rm yr}^{-1}$ \citep{2021arXiv211103634T}. 
The non-detection has also been used to put empirical constraints on the EM counterpart event rate and luminosity function. The non-detection of kilonovae over a period of $35$ months of ZTF data provides an upper limit to the kilonova event rate $<900\,{\rm Gpc}^{-3}{\rm yr}^{-1}$ \citep{Andreoni20216363}. The non-detection of EM counterpart during the LVC's O3 run suggests that $\lesssim80\%$ (or $\lesssim10\%$) of kilonovae are brighter than $\sim-16.0$\,mag (or $\sim-17.0$\,mag) \citep{Kasliwal2020145145}.

It is expected there will be many more BNS merger events with the improving sensitivities of the LIGO/VIRGO/KAGRA (LVK) detectors and the planned third-generation ground-based GW detectors, such as Einstein Telescope (ET) and Cosmic Explorer (CE) \citep[e.g.,][]{Reitze20193535, Maggiore20205050}. The EM counterparts of some of those events with small distances will be eventually detected \citep[e.g.,][]{Scolnic-2018-3-3, Setzer-2019-4260-4273} and the luminosity function of these events may also be obtained. Different from the properties of individual events, the luminosity function, as a statistical property, may also help to put additional constraint on the physical processes related to the EM counterparts and reveal the nature of NSs.

The EM phenomena associated with a BNS merger are the short gamma-ray burst (sGRB) and kilonova \citep{Jin2017, Troja-2017-71-74, Troja201921042116, Savchenko20171515, Kim20172121, Nakar20181818, Granot201815971608, Ascenzi2019672690, Rossi202033793397}, which are produced by the jet launched during the merger \citep{Granot201815971608, Ioka2018243, Nakar20181818} and mainly the ejecta from the merger \citep[e.g.,][]{Metzger201026502662, Pian-2017-67-70, Wanajo20186565, Metzger-2019-1-1, Siegel2019203203}, respectively. The main energy source of EM radiation from a kilonova is the r-process decay of heavy nuclei (e.g., lanthanide and actinide elements) occurring in the ejecta from  mergers. The ejecta include the ``dynamical ejecta'' immediately released after the merger, as well as the ``post-merger outflow'' from the remnant disc on a time-scale of seconds. Therefore, the light curves (LCs) of a kilonova is mainly determined by the masses of both ejecta components and their dynamical evolution, which are controlled by the total mass, mass ratio, and the equation of state (EOS) of the BNS \citep{Kawaguchi-2020-171-171}.  The main goal of this paper is to study the dependence of the kilonova properties on the BNS physical parameters and the EOS of NSs, model the kilonova luminosity functions (KLFs) systematically, and investigate whether the KLF depends on the EOS or not and thus whether it can also give a statistical constraint on the EOS. To do this, we need a detailed model for kilonova, the properties of the ejecta from BNS mergers, and also the distributions of BNS properties among the BNS merger population under different EOSs.

Different approaches have been developed for understanding of the kilonova phenomenon. For example, the semi-analytical models for the kilonova radiation processes have been intensively studied in the past years, to either predict the kilonova LCs or interpret the observed kilonova phenomenon \citep[e.g.,][]{Li-1998-59-62, Perego-2017-37-37, Villar-2017-21-21, Metzger-2019-1-1}. The full general relativistic numerical (NR) simulations have been developed and performed to estimate the properties of the dynamical ejecta from and the remnant disc of BNS and NS-black hole (NSBH) mergers \citep[e.g.,][]{Dietrich-2017-105014-105014,Fahlman-2018-3-3,Radice201836703682,Fernandez-2019-4-16,Nedora20219898}, which drives the kilonova phenomenon. More sophisticated simulations were also developed to directly predict the kilonova LCs and spectra by the combination of the radiative transfer and NR simulations \citep[e.g.,][]{Kawaguchi-2020-171-171, Korobkin2021116116, Wollaeger20211010, Pang202222058513}. As there will be many kilonovae to be detected in the near future with detailed LCs and spectra, the latter approach may be urgently required for detailed understanding of the kilonova phenomena and the BNS/NSBH mergers themselves. However, it is still a challenge to do a large number of such simulations because of computation consuming. 

In this paper, we adopt a semi-analytical model to predict the kilonova LCs from different BNS mergers with different physical parameters under different EOSs, which is based on the general framework developed by many authors \citep[e.g.,][]{ Perego-2017-37-37, Villar-2017-21-21, Metzger-2019-1-1}. 
The simple semi-analytical model may be easier and more efficient to be applied for investigating large parameter space, especially when combining with the dynamical ejecta properties predicted by NR numerical simulations, which is required for investigating the population properties (e.g., luminosity functions) of kilonovae.
We also adopt the predictions on the ejecta properties from BNS mergers with different total mass, mass ratio, and EOS given by NR simulations. We estimate the distribution functions of kilonova peak luminosities at different bands by the combination of the kilonovae model, the population synthesis model for BNS formation in \citet{Chu202215571586}. 

The paper is organized as follows. In Section~\ref{sec:model}, we overview the semi-analytical model, as a general frame work, for modelling kilonova. In Section~\ref{sec:GW170817}, we adopt the semi-analytical kilonova model (implemented into the \texttt{MOSFIT} code) to fit the LCs of GW170817 and obtain the constraints on all the physical model parameters. In Section~\ref{sec:gkilonova}, we estimate the LCs for a general kilonova with a given set of total mass and mass ratio under any given EOS, by using the properties of the ejecta inferred from GW numerical simulations and other physical parameters inferred from the model fitting to the GW170817 LCs. We analyse the dependence of the multiband radiation from the kilonova (especially the peak magnitude of the LCs) on the total mass and mass ratio, etc.  We also estimate the LCs of GW190425, the second BNS merger discovered by the LIGO and Virgo. In Section~\ref{sec:LF}, we estimate the KLF by incooperating the BNS merger rate given by simple BNS formation models. We analyse the dependence of the KLF on the adopted EOS. Conclusions are summarized in Section~\ref{sec:conclusion}.

\section{Kilonova model}
\label{sec:model}

The LCs of a kilonova depend on the properties (mass, velocity, and opacity) of the ejecta from a BNS (NSBH) merger via the dynamical process and winds, since the main energy source is the radioactive decay of r-process heavy elements in the ejecta during the merging process. The heating from the fallback disc to the ejecta may also contribute some to the emergent radiation and thus affects kilonova LCs. Below we introduce the general framework for estimating kilonova LCs for an arbitrary BNS merger.

\subsection{Dynamical Ejecta}
\label{subsec:dyn_ej}

When a BNS merge, a small fraction of its neutron matter ($\sim 10^{-4}-10^{-2} M_{\odot}$) can be ejected out on the dynamical time-scale, denoting as the ``dynamical ejecta'', with a velocity of $\sim 0.1-0.3c$ \citep[e.g.,][]{Metzger-2019-1-1}. Numerical relativity simulations have shown that the dynamical ejecta component close to the polar direction has low opacity (hereafter denoting as the dynamical blue (db) component), while that close to the equatorial plane has relatively higher opacity (hereafter denoting as the dynamical red (dr) component) \citep[e.g.,][]{Radice-2018-130-130}. \citet{Perego-2017-37-37} and \citet{Radice-2018-130-130} developed semi-analytical models based on the results of NR simulations for kilonova. We refer to these models, and set a model parameter $\theta_{\rm db}$ as the half-opening angle of the db component, so that matter in the direction of $\theta \in [0^{\circ},\theta_{\rm db}]$ belongs to the db component, and that with $\theta \in (\theta_{\rm db},90^{\circ}]$ represents the dr component. The mass of dynamical ejecta is assumed to be uniformly distributed over the solid angle except that the opacity for the db and dr components are different, as NR simulations suggested \citep[e.g.,][]{Perego-2017-37-37, Radice-2018-130-130}. Then we may define the equivalent mass for both red ($m_{\rm eq}^{\rm dr}$) and blue ($m_{\rm eq}^{\rm db}$) components as the one equal to the total mass of dynamical ejecta $m_{\rm ej}^{\rm d}$, i.e., 
\begin{equation}
m_{\rm eq}^{\rm dr} = m_{\rm eq}^{\rm db} = m_{\rm ej}^{\rm d}.
\end{equation}
According to this equivalent mass we can obtain the equivalent luminosity for both the dr and db component as to be described in Section~\ref{subsec:LCs}.

\subsection{Disc Wind} 
\label{subsec:wind}

Outflows from the remnant disc of compact binary mergers, occurring over a time-scale of seconds or longer, can also contribute significantly to the ejecta mass \citep[e.g.,][]{Metzger-2019-1-1}. Outflows may be generated from the disc either via the neutrino- or viscous-driven processes. We link the disc mass $m_{\rm disc}$ with the neutrino- and viscous-driven disc wind mass by the proportion factors $\xi_{\nu}$ and $\xi_{\rm vis}$, respectively, which are chosen as our model parameters. There are significant distinctions in the opacity, velocity, and density between these two kinds of outflows.

For the neutrino-driven wind, we assume that the ejecta mass is uniformly distributed over $\cos\theta$ with $\theta \in [0^{\circ}, \theta^{\rm \nu p}]$ and there is no wind in the direction with $\theta> \theta^{\nu \rm p}$. The opacity is assumed to be $\kappa= \kappa_{\rm b}$ when $\theta \in [0^\circ, \theta^{\rm \nu b}]$, and the electron fraction ($Y_{\rm e}$) of the post-merger ejecta is found to be high in NR simulations (hereafter denoted as the neutrino-driven wind blue ($\nu$b) component) \citep[e.g.,][]{Radice-2018-130-130}. The opacity of the rest part with $\theta \in (\theta^{\rm \nu b}, \theta^{\rm \nu p})$ is assumed to be $\kappa =\kappa_{\rm p}$, where a moderate average $\langle Y_{\rm e}\rangle$ is expected (hereafter denoted as the neutrino-driven wind purple ($\nu$p) component) \citep[e.g.,][]{Radice-2018-130-130}. The opening angles of the $\nu$b and $\nu$p components of the neutrino-driven disc wind $\theta^{\rm \nu b}$ and $\theta^{\rm \nu p}$, and its opacity $\kappa_{\rm b}$ and $\kappa_{\rm p}$, are taken to be constant model parameters, respectively. Thus, the masses of the $\nu$b and $\nu$p components are
\begin{equation}
m_{\rm ej}^{\rm \nu  b}=\left(\frac{1-\cos{\theta^{\rm \nu b}}}{1-\cos{\theta^{\rm \nu p}}}\right) m_{\rm ej}^{\nu},
\end{equation}
and
\begin{equation}
m_{\rm ej}^{\nu\rm p}=\left(\frac{\cos{\theta^{\rm \nu  b}}-\cos{\theta^{\rm \nu  p}}}{1-\cos{\theta^{\rm \nu p}}}\right)m_{\rm ej}^{\nu},
\end{equation}
respectively, where $m^\nu_{\rm ej}$ is the total mass ejected by the neutrino-driven disc wind.
Consequently, the equivalent mass for the $\nu$b and $\nu$p components of the neutrino-driven disc wind can be defined  the same way as that for the db and dr components of the dynamical ejecta (see Section~\ref{subsec:dyn_ej}), i.e.,
\begin{equation}
m_{\rm eq}^{\rm \nu b}=\frac{m_{\rm ej}^{\rm \nu b}}{1-\cos{\theta^{\rm \nu b}}},
\end{equation}
and
\begin{equation}
m_{\rm eq}^{\rm \nu p}=\frac{m_{\rm ej}^{\rm \nu  p}}{\cos{\theta^{\rm \nu b}}-\cos{\theta^{\rm \nu p}}}.
\end{equation}

For the viscous-driven disc wind, we assume the mass distribution is $\propto \sin^2 \theta$ according to \citet{Perego-2017-37-37}. We evenly divide the cosine of azimuthal angle $\cos\theta$ into $i=1,2,\cdots,5$ slices. In the interval with $\cos\theta\in [\cos\theta_i, \cos\theta_{i+1}]$, the corresponding equivalent ejecta mass is 
\begin{equation}
m_{i,\rm eq}^{\rm vis}= \frac{1}{2}m_{\rm ej}^{\rm vis}\left(3-\cos^2{\theta_i}-\cos^2{\theta_{i+1}}-\cos{\theta_i}\cos{\theta_{i+1}}\right),
\end{equation}
where $m_{\rm ej}^{\rm vis}$ is the total mass ejected out via the viscous-driven disc wind. For the first and the last slices of this viscous-driven wind (denoted as the visp component), we consider a purple opacity, whereas the rest slices (denoted as the visr component) are assumed with the red opacity. The equivalent luminosity from the viscous-driven disc wind can be then obtained for each slice.

\subsection{Kilonova luminosity}
\label{subsec:luminosity}

Transients from BNS or BH-NS mergers are expected to be mainly powered by the radioactive decay of r-process elements taking places in the neutron-rich ejecta \citep[e.g.,][]{Li-1998-59-62, Villar-2017-21-21, Perego-2017-37-37, Metzger-2019-1-1, Zhu-2020-20-20}. 
In this paper, we adopt a simple model described below to generate LCs for kilonovae associated with BNS mergers as in \citet{Perego-2017-37-37}, i.e., 
\begin{equation}
L_{\rm in}(t) =  {\epsilon_0} {\epsilon_{\rm Y_{\rm e}}(t)} \left(\frac{\epsilon_{\rm th}(t)}{0.5}\right) M_{\rm ej}\left[\frac{1}{2}-{{\frac{1}{\pi}}\arctan\left({\frac{t-{t_0}}{\Delta t}}\right)}\right]^{1.3}.
\label{eq:inlum} 
\end{equation}
Here $t_0=1.3$\,s, $\Delta t=0.11$\,s, $M_{\rm ej}$ is the ejecta mass, $\epsilon_0$ is a model parameter with $2\times10^{18} {\rm erg\,s^{-1}\,g^{-1}} \lesssim \epsilon_0 \lesssim 2\times10^{19} {\rm erg\,s^{-1}\,g^{-1}}$ because of the large uncertainties of the nuclear mass and decay models, $\epsilon_{\rm th}$ is the thermalization efficiency described as \citep{Barnes-2016-110-110}
\begin{equation}
\epsilon_{\rm th}(t)=0.36\left[\exp\left(\frac{0.56t}{1\rm day}\right)+\frac{\ln\left(1+0.34(t/1{\rm day})^{0.74}\right)}{0.34(t/{1\rm day})^{0.74}}\right],
\end{equation}
and $\epsilon_{\rm Y_{\rm e}}(t)$ is a factor to denoting the significant heating difference for neutron-rich ejecta with electron-fraction $Y_{\rm e}\geq 0.25$, i.e.,
\begin{equation}
\epsilon_{\rm Y_{\rm e}}(t)= \\
\left\{
\begin{aligned}
\frac{1}{2} + \frac{\epsilon_{\rm n}}{{  1 + \exp\left[\epsilon_{\rm dec}\left(\frac{t-t_{\rm n}}{1{\rm day}}\right)\right] }}, & \ \ \ \textrm{if } Y_{\rm e} \geq 0.25, \\
1, & \ \ \ \textrm{otherwise},
\end{aligned} \right.
\label{eq:free neu}
\end{equation}
where $\epsilon_{\rm n}$, $\epsilon_{\rm dec}$, and $t_{\rm n}$ are parameters that can be constrained by the LCs of  AT2017gfo. For each ejecta component, its equivalent luminosity can be calculated according to the above Equation~\eqref{eq:inlum}, with $M_{\rm ej}$ describing the equivalent mass of that ejecta component.

\subsubsection{Heating from fallback accretion}
\label{subsec:fallback}

In addition to the neutrino-driven and viscous-driven winds, there is an extra wind powered by the fallback accretion \citep{Decoene-2020-45-45}, that is, the accretion power radiated from the fallback disc heats a part of material, throws it out and boosts the kilonova luminosity \citep{Metzger-2019-1-1}. Therefore, we also incorporate the fallback accretion as an additional energy supply to yield a separate radiation component (as constrained by GW170817 observations, see \citealt{Wollaeger-2019-22-22}). The fallback accretion power $\dot{Q}_{\rm fb}(t)$ is assumed to follow the canonical power law as \citep[e.g.,][]{Metzger-2019-1-1, Wollaeger-2019-22-22, Decoene-2020-45-45, Ishizaki-2021-13-13}
\begin{equation}
\begin{split}
\dot{Q}_{\rm fb}= 2\times 10^{51} {\rm erg\,s^{-1}} \epsilon_j 
\left[\frac{\dot{M}_{\rm fb}(0.1{\rm s})}{10^{-3}M_\odot}\right]
\left(\frac{t}{0.1{\rm s}}\right)^{-5/3}.
\label{eq:fb}
\end{split}
\end{equation}
Here $\epsilon_j$ is the mass-to-energy conversion efficiency, assuming the canonical value of $0.1$, and $\dot{M}_{\rm fb}(0.1{\rm s})$ is the fallback rate at $t=0.1$\,s. In our model, the mass of the layer, in which the heating from the fallback accretion diffuses and radiates out, is denoted as $m_{\rm fb,ej}$, the velocity of the associated photosphere is denoted as $v_{\rm fb}$, and these two parameters are assumed to be constant. We assume that this fallback accretion induced radiation component is  isotropic.

\subsubsection{Emergent luminosity}
\label{sec:diffusion}

Photons are reprocessed when they cross the ejecta because of significant opacity, and the emergent radiation may be simply approximated as the blackbody radiation. Therefore, the observational luminosity should be different from the intrinsic one computed from Equation~\eqref{eq:inlum} or \eqref{eq:fb}. As that in most semi-analytical kilonova models \citep[e.g.,][]{Villar-2017-21-21,Perego-2017-37-37}, we also use the implementation of \citet{Arnett-1982-785-797} to estimate the emergent luminosity as
\begin{equation}
L_{\rm emerg}\left(t\right) = \exp{\left(-\frac{t^2}{t_{\rm d}^2}\right)}
 \int_{0}^{t}\exp{\left(-\frac{{t^\prime}^2}{t_{\rm d}^2}\right)}\left(\frac{t^\prime}{t_{\rm d}}\right)L_{\rm in}\left(t^\prime\right)d \left(\frac{t^\prime}{t_{\rm d}} \right) , \\
\label{eq:obs_lum}
\end{equation}
where $L_{\rm in}$ is the input luminosity from the central heating source, $t_{\rm d}=\sqrt{2 \kappa M_{\rm ej}/{\beta}cv_{\rm sc}}$ is the diffusion time-scale characterizing the time when photons escape the ejecta faster than that the ejecta can expand, with $\kappa$ representing the opacity, $\beta$ ($=13.4$) representing a dimensionless constant about the geometry of the ejecta, and $v_{\rm sc}$ the expanding velocity of the ejecta  \cite[]{Arnett-1982-785-797, Chatzopoulos-2012-121-121}.

\subsubsection{Photosphere of the ejecta}
\label{sec:photosphere}

The temperature of an expanding spherical photosphere with luminosity $L_{\rm emerg}(t)$ via blackbody radiation can be simply estimated as \citep[e.g.,][]{Nicholl-2017-55-55}
\begin{equation}
T_{\rm ph}(t)= \left\{
\begin{aligned}
&\left[\frac{L_{\rm emerg}\left(t\right)}{4\pi{\sigma_{\rm SB}}v_{\rm ph}^2t^2}\right]^\frac{1}{4},& \ \ \  \textrm{if } t< t_{\rm c} & \\
&T_{\rm f},& \ \ \ \textrm{if } t\geq t_{\rm c},&\\
\end{aligned} \right.
\label{eq:Tph}
\end{equation}
where $\sigma_{\rm SB}$ is the Stefan-Boltzmann constant, $v_{\rm ph}$ is the photosphere expansion velocity and assumed to be a constant, and $t$ is the elapsed time. For different components of a kilonova, the temperatures of the associated photospheres can be significantly different.

The radius of the photosphere of an ejecta component is 
\begin{equation}
R_{\rm ph}(t)=
\left\{
\begin{aligned}
& v_{\rm ph}t,  \ \ \ & \textrm{if } t< t_{\rm c}, \\
&\bigg[\frac{L_{\rm emerg}\left(t\right)}{4\pi{\sigma_{\rm SB}}{T_{\rm f}}^4}\bigg]^\frac{1}{2},  & \ \ \ \textrm{if } t\geq t_{\rm c},	
\end{aligned} \right.
\label{eq:Rph}
\end{equation}
where $t_{\rm c} = \left[ L_{\rm in} \left(t\right)/4\pi{\sigma_{\rm SB}}v_{\rm ph}^2 T_{\rm f}^4\right]^{1/2}$ marks the time when $T_{\rm ph}(t)$ becomes flat ($=T_{\rm f}$) when the elapsed time $t\geq t_{\rm c}$ and the photosphere begins receding after it reaches the maximum value $v_{\rm ph} t_{\rm c}$. 

In the above model, the temperature of the ejecta is simply assumed to be the effective one derived from the Stefan-Boltzmann law during the homologous expansion stage, and after that the ejecta cools down to a stage with a constant temperature $T_{\rm f}$. The late-time shapes of LCs are directly related to the final temperature floor $T_{\rm f}$ \citep[][]{Nicholl-2017-55-55}. Throughout this paper, we adopt $T_{\rm f}$ as a model parameter (see Table~\ref{tab:t1}).

\subsubsection{Kilonova light curves}
\label{subsec:LCs}

The luminosity of blackbody radiation from a spherical photosphere at the wavelength $\lambda_{\rm e} \rightarrow \lambda_{\rm e} + d\lambda_{\rm e}$ in the rest frame is given by
\begin{equation}
L_{\lambda_{\rm e}}=4{\pi}R_{\rm ph}^2B_{\lambda_{\rm e}}=\frac{8{\pi}^2R_{\rm ph}^2 h c^2/{\lambda_{\rm e}^5}}{\exp(hc/k\lambda_{\rm e} T_{\rm ph})-1},
\label{eq:sph_blackbody}
\end{equation}
where $\lambda_{\rm e}$ represents the wavelength of a photon emitted at the photosphere and it is related to the photon wavelength at the observer's rest frame ($\lambda_{\rm obs}$) as $\lambda_{\rm e}={\lambda_{\rm obs}}/(1+z)$. We can use the above equation to obtain the equivalent luminosity for each components described above for any kilonova. 

The actual flux of a kilonova received by a distant observer, however, depends on its viewing angle since the photosphere of each component is normally not spherical. Here, we adopt the recipe same as that given in \citet{Martin-2015-2-2} to estimate this dependence as following. The full solid angle of $4\pi$ is divided into many solid angle elements with $\boldsymbol{n}_k$ describing the unit vector normal to the $k$-th element of the photosphere and $\boldsymbol{q}$ representing the unit vector pointing along the viewing direction of the observer. The viewing angle ($\theta_{\rm v}$) is defined as the angle between the viewing direction of the observer and the normal direction of the disc. Then the geometric projection factor of the photosphere for each element (covering a part of the sky of the central source) on to the observer's sky plane is given by
\begin{equation}
p_k=\frac{1}{\pi}\iint_{\boldsymbol{n}_k \cdot \boldsymbol{q}>0}\boldsymbol{q}\cdot d\boldsymbol{\Omega}.
\label{eq:view angle}
\end{equation}
Here, the condition $\boldsymbol{n}_k \cdot \boldsymbol{q}>0$ is required as only the part of the surface facing to the observer can be visible. Different ejecta components have different projection factors for a fixed $\boldsymbol{q}$. For each observed kilonova, such as GW170817, its viewing angle may be directly constrained by fitting its LCs.

For the photosphere of each kilonova component, the geometric projection factor can be obtained by summing up $p_k$ over all the solid angle elements within that specific photosphere. We denote these geometric projection factors as $p^{\rm db}$, $p^{\rm dr}$, $p^{\rm \nu b}$, $p^{\nu \rm p}$, $p^{\rm visp}$, and $p^{\rm visr}$ for the db and dr components of the dynamical ejecta, the $\nu$b and $\nu$p components of the neutrino-driven disc wind, and  purple and red components of the viscosity-driven disc wind, respectively. Here $p^{\rm visp}$ is obtained by summing up those for the first and second slices of the viscous-driven wind, while $p^{\rm visr}$ is obtained by summing up those for the other three slices of the wind (see Section~\ref{subsec:wind}). For the assumed spherical photosphere corresponding to the heating from the fallback accretion, the geometric projection factor is $1$. Therefore, the luminosity of a kilonova measured by a distant observer with a viewing angle $\theta_{\rm v}$ should be the summation of the equivalent luminosity of each component multiplied by the geometric projection factor, i.e., 
\begin{equation}
L^{\rm tot}_{\lambda_{\rm E}}(t)=\sum_j p^j L_{\lambda_{\rm E},j},
\end{equation}
where $j$ denotes the $j$-th component of the kilonova, including all those mentioned above, e.g., $p^{\rm db}$, $p^{\rm dr}$, $p^{\rm \nu b}$, $p^{\nu \rm p}$, $p^{\rm visp}$, and $p^{\rm visr}$, $L_{\lambda_{\rm E},j}$ represents the equivalent luminosity calculated by Equation~\eqref{eq:sph_blackbody} with $T_{\rm ph}$ and $R_{\rm ph}$ given by Equations~\ref{eq:Tph} and \eqref{eq:Rph} for each component, respectively.

For any given filter \textbf{X}, the AB magnitude for a kilonova at a given time $t$ can be then directly obtained by using its spectral energy distribution (SED) and the response function of that filter ($S(\lambda)$) as \citep[see also][]{Nicholl-2017-55-55}
\begin{equation}
m_{\textbf{X}}(t)=m_0-2.5\log (L_{\rm E}(t_{\rm obs})/\textrm{erg}~\textrm{s}^{\rm -1}) +5\log(d_{\rm L}/\textrm{Mpc})
\label{eq:magX}
\end{equation}
where $m_0$ zero point, the effective luminosity $L_{\rm E}$ is given by
\begin{equation}
L_{\rm E}(t_{\rm obs})=\frac{\int S\left(\lambda_{\rm obs}\right) L^{\rm tot}_{\lambda_{\rm obs}/(1+z)}(\frac{t_{\rm obs}}{1+z})/(1+z) d\lambda_{\rm obs}}{\int S\left(\lambda_{\rm obs}\right) d\lambda_{\rm obs}},
\label{eq:LE}
\end{equation}
and $d_{\rm L}$ is the luminosity distance of the kilonova. We also consider the time dilation due to the cosmic expansion. All the time quantities in Equations~\eqref{eq:inlum} - \eqref{eq:Rph} denote the time in the kilonova rest frame.

For any kilonova with multiband observations, one may adopt the above model to constrain the mass, velocity, and other properties of each ejecta component by fitting the LCs. Currently, only two BNS mergers, GW170817 and GW190425, are detected by LIGO/Virgo, and only the EM counterpart of GW170817, AT2017gfo, has been intensively observed. Below we adopt the kilonova model introduced above to fit the AT2017gfo LCs and obtained the best-fitting values of the model parameters, including the dynamical ejecta mass $m^{\rm d}_{\rm ej}$, the disc mass $m_{\rm disc}$, the velocity $v_{\rm dyn}$, opening angle of the blue component of the dynamical ejecta $\theta_{\rm db}$, the decay of free neutrons in the outer layer ($\epsilon_{\rm n}$, $\epsilon_{\rm dec}$, $t_{\rm n}$ in equation~\ref{eq:free neu}), the coefficient of the r-process heating power $\epsilon_{\rm 0}$, the fallback mass accretion rate $\dot{M}_{\rm fb}(0.1{\rm s})$, the mass of the layer through which the fallback accretion heating energy diffuses $m_{\rm ph}^{\rm fb}$, the fractions of disc mass ejected in the form of neutrino-driven wind and viscous-driven wind ($\xi_{\nu}$, $\xi_{\rm vis}$), an additional deviation $\sigma$ that is added to all measurement errors encompassing additional source of ``white noise'', the opening angle of the blue and purple components of the neutrino-driven wind ($\theta^{\rm {\nu}b}$, and $\theta^{\rm {\nu}p}$), the viewing angle of the observer $\theta_{\rm v}$, the opacities of the blue, purple, red, and fallback photosphere components ($\kappa_{\rm b}$, $\kappa_{\rm p}$, $\kappa_{\rm r}$ and $\kappa_{\rm fb}$), the photosphere temperature floor of the blue, purple, red, and fallback components ($T_{\rm b}$, $T_{\rm p}$, $T_{\rm r}$ and $T_{\rm fb}$), the velocity of the photosphere layer of fallback accretion $v_{\rm fb}$, the neutrino-driven wind velocities of the blue and purple components ($v^{\rm {\nu}b}$ and $v^{\rm {\nu}p}$), the viscous wind velocities of purple and red components ($v^{\rm {visp}}$ and $v^{\rm {visr}}$). Note that we assume that the temperatures/opacities of the db and $\nu$b components are the same, and the temperatures/opacities of the dr and visr components are also the same. For the velocities of these different components, we assume that both the blue and red components of the dynamical ejecta have the same velocity $v_{\rm dyn}$ for simplicity as NR simulations suggest a narrow velocity distribution of dynamical ejecta \citep{Dietrich-2017-105014-105014, Fernandez-2019-4-16}. However, we assume that different components of other ejecta have different velocities since NR simulations indicate the viscous disc wind velocity show a broader distribution than that of the dynamical ejecta \citep{Fernandez-2019-4-16}. 

\section{Light curves of GW170817 and model fitting}
\label{sec:GW170817}
\subsection{Observations of the kilonova associated with GW170817}
\label{subsec:Obs_GW170817}

As the first BNS merger with both GW and EM signal detections, GW170817/AT2017gfo has abundant EM measurements in the UV-optical-infrared (UVOIR) bands \citep[e.g.,][]{Andreoni-2017-69-69, Arcavi-2017-64-66, Coulter-2017-1556-1558, Cowperthwaite-2017-17-17, Diaz-2017-29-29, Drout-2017-1570-1574, Evans-2017-1565-1570, Hu-2017-1433-1438, Valenti-2017-24-24a, Kasliwal-2017-1559-1565, Lipunov-2017-1-1, Pian-2017-67-70, Pozanenko-2018-30-30, Shappee-2017-1574-1578, Smartt-2017-75-79, Tanvir-2017-27-27, Troja-2017-71-74, Utsumi-2017-101-101}. We adopt the pruned and homogenized UVOIR data set for GW170817/AT2017gfo 
given in \citet[][see table therein]{Villar-2017-21-21}, of which the systematic offsets between different data are corrected. 
We adopt the \texttt{MOSFIT} code introduced below to constrain the model parameters described above in Section~\ref{sec:model} by fitting this UVOIR data set of GW170817/GT2017gfo.

\subsection{MOSFIT}
\label{subsec:MOSFIT}
\texttt{MOSFIT} is a Python code specialized for fitting the LCs of transients with various semi-analytical models using encapsulated modules for different input energy sources, diffusion methods, prescriptions for photospheres, SEDs, fitting routines, etc. (see details in \citealt{Guillochon-2018-6-6}). In particular, we implemented the disc wind model in the \citet{Perego-2017-37-37} with a little modification about the opacities and velocities of disc wind according to the recent NR simulation results as described in Section~\ref{subsec:wind}. We further independently add extra engines, including the free neutron heating of outer ejecta (see equation~\ref{eq:free neu}) and the fallback accretion heating (see equation~\ref{eq:fb}), into our model. The projection effect of the fluxes into the observing direction (equation~\ref{eq:view angle}) is coded into the SED module.

\texttt{MOSFIT} builds a call stack stipulating the executing sequence of each relevant module. When the fitting begins, an \texttt{emcee} sampler \citep[][]{ForemanMackey-2013-306-306, ForemanMackey-2019-1864-1864} is called and generated by \texttt{MOSFIT}, this sampler draws walkers from the priors of model parameters and iterates the `Burning' and the `Walking' stages attempting to achieve the optimized parameter distributions. A successful convergence of fitting is reached when the potential scale reduction factor (PSRF, also known as the Gelman–Rubin statistic; \citealt[][]{Gelman-1992-457-472}) is lower than a specified threshold (e.g., ${\rm PSRF} < 1.1$ by default). The performance of the fitting results are quantified by the scores of the Watanabe-Akaike information criterion (WAIC). In order to improve the fitting efficiency, we apply the multicore parallelization technique for our fitting of the UVOIR data set of GW170817/AT2017gfo. There are totally $29$ model parameters to describe those various physical processes in generating the kilonova LCs. We adopt these $29$ parameters as free model parameters when using the \texttt{MOSFIT} mode to fit the LCs of GW170817 (see Table~\ref{tab:t1}). 

\subsection{Fitting results}
\label{subsec:GW170817fit}

\begin{figure}
\centering
\includegraphics[scale=0.30]{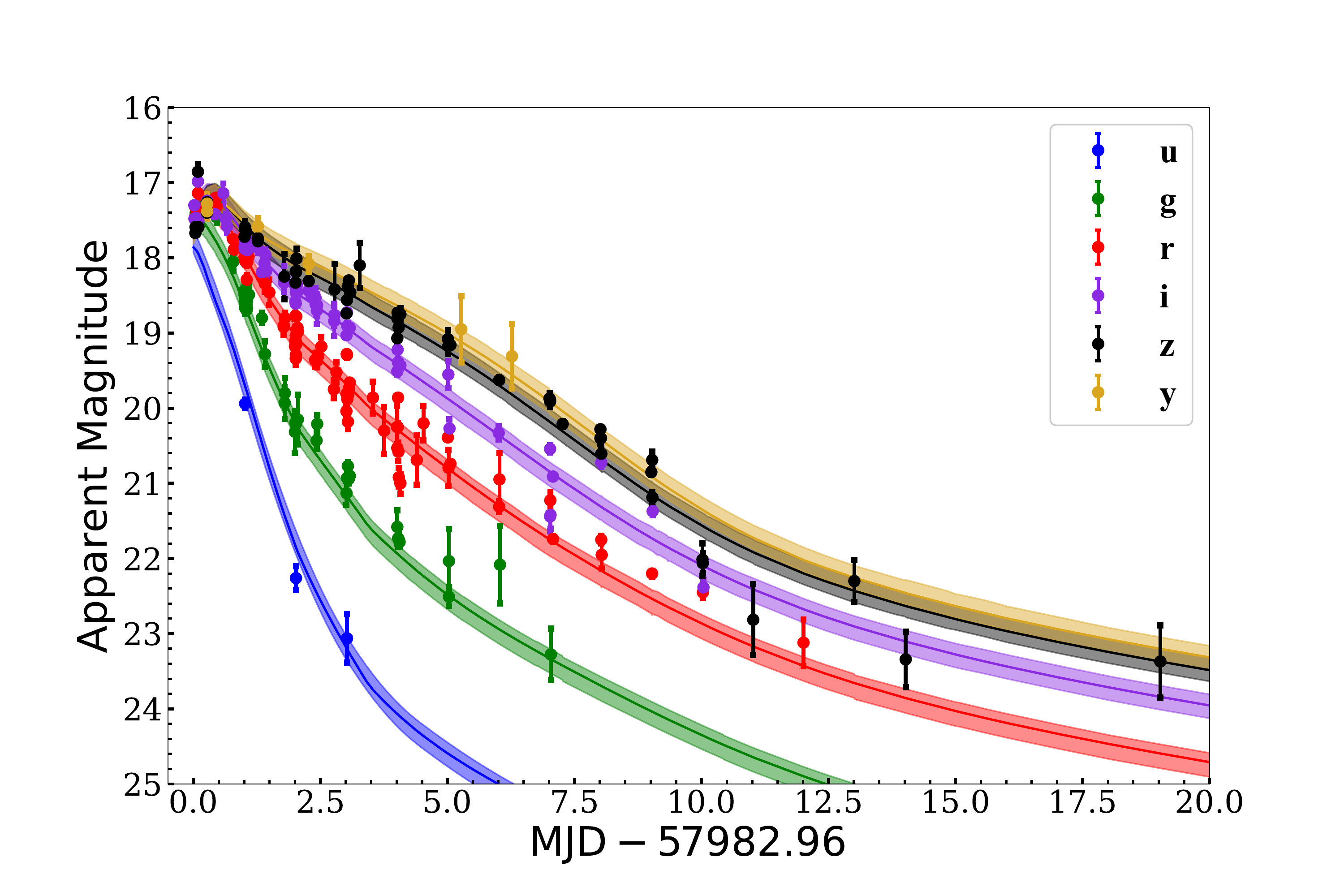}
\caption{
Multiband LCs of GW170817 and the best model fit. Blue, green, red, purple, black, and yellow symbols represent the measurements of the u, g, r, i, z, and y bands, respectively, and the solid lines with corresponding colours represent the best fits to the u, g, r, i, z, and y band LCs, with the associated shaded areas representing the $1\sigma$ uncertainties of the model fits. 
}
\label{fig:f1}
\end{figure}

We obtain the best fit to the LCs of GW170817 after running $\gtrsim30000$ MCMC iterations with $\gtrsim1000$ walkers using \texttt{MOSFIT} ($9000$ of iterations being `burning' and the rest of them running `walking') with a final WAIC mark of $890$, which shows our model is acceptable. 

Figure~\ref{fig:f1} shows the best fit to the LCs of GW170817 and Table~\ref{tab:t1} lists the best-fitting values of those $29$ model parameters and their uncertainties. 
The two-dimensional and one-dimensional probability distributions of the best fits to the model parameters are shown in Figures~\ref{fig:fa1}, \ref{fig:fa2}, and \ref{fig:fa3} in the appendix.
The best-fitting model suggests that the disc mass, the dynamical ejecta mass and velocity, and the opening angle for the blue component of the dynamical ejecta are $m_{\rm disc}\sim 0.11M_\odot$, $m^{\rm d}_{\rm ej} \sim 0.0038M_\odot$, $v_{\rm dyn}\sim 0.285 c$, and $\theta_{\rm bd} \sim 30.8^\circ$, respectively. As seen below, the best-fitting values of these model parameters lie in the physically acceptable ranges, and they are compatible with the available results obtained from the NR dynamical simulations and the constraints from GW170817 observations.

In the literature, there are a number of NR dynamical simulations performed for BNS mergers similar to GW170817 and analysed their ejecta properties. For example, \cite{Nedora20219898} performed NR dynamical simulations for $37$ GW170817-like BNS mergers, and find $m_{\rm disc} \sim 0.1-0.3M_{\odot}$,  $m^{\rm d}_{\rm ej} \sim10^{-3}-10^{-2}M_{\odot}$, $v_{\rm dyn} \sim0.1-0.3c$, and $\theta_{\rm bd} \sim20-40^{\circ}$, depending on the exact value of the BNS mass ratio [$q\in(1,1.8)$]. Apparently, our best-fitting values for these several parameters are roughly consistent with the simulation results, though the best-fitting value for $v_{\rm dyn}$ seems to be close to the upper bound of the simulation ones  (see Table~\ref{tab:t1}). \citet{Radice201836703682} also reported the simulation results for the merger of a $1.35M_{\odot}+1.35M_{\odot}$ BNS with the EOS of DD2 \citep[c.f.][]{Typel20101580315803}. They found that the neutrino-driven wind should be $m^{\nu}_{\rm ej} \sim10^{-3}M_{\odot}$ while the viscous-driven wind  $m_{\rm ej}^{\rm vis}\sim0.05 - 0.2M_{\odot}$ (depending on whether including neutrino absorption or not). Our fitting results ($m^{\nu}_{\rm ej} = \xi_\nu m_{\rm disc} \sim 0.0015 M_\odot$, $m^{\rm vis}_{\rm ej} = \xi_{\rm vis} m_{\rm disc} \sim 0.029 M_\odot$; see Table~\ref{tab:t1}) are also roughly consistent with their simulations, though the best-fitting value of $0.03 M_\odot$ is close to the lower bound of the simulation ones. We also note that the total ejecta mass ($\sim 0.037M_\odot$, including the $m^{\rm fb}_{\rm ph}$) given by the best fit is consistent with those obtained by \citet[][$\sim 0.06M_\odot$]{Perego-2017-37-37} and \citet[][$\sim 0.01-0.08M_\odot$]{Setzer-2019-4260-4273}. 

\begin{figure*}
\includegraphics[width=0.8\textwidth]{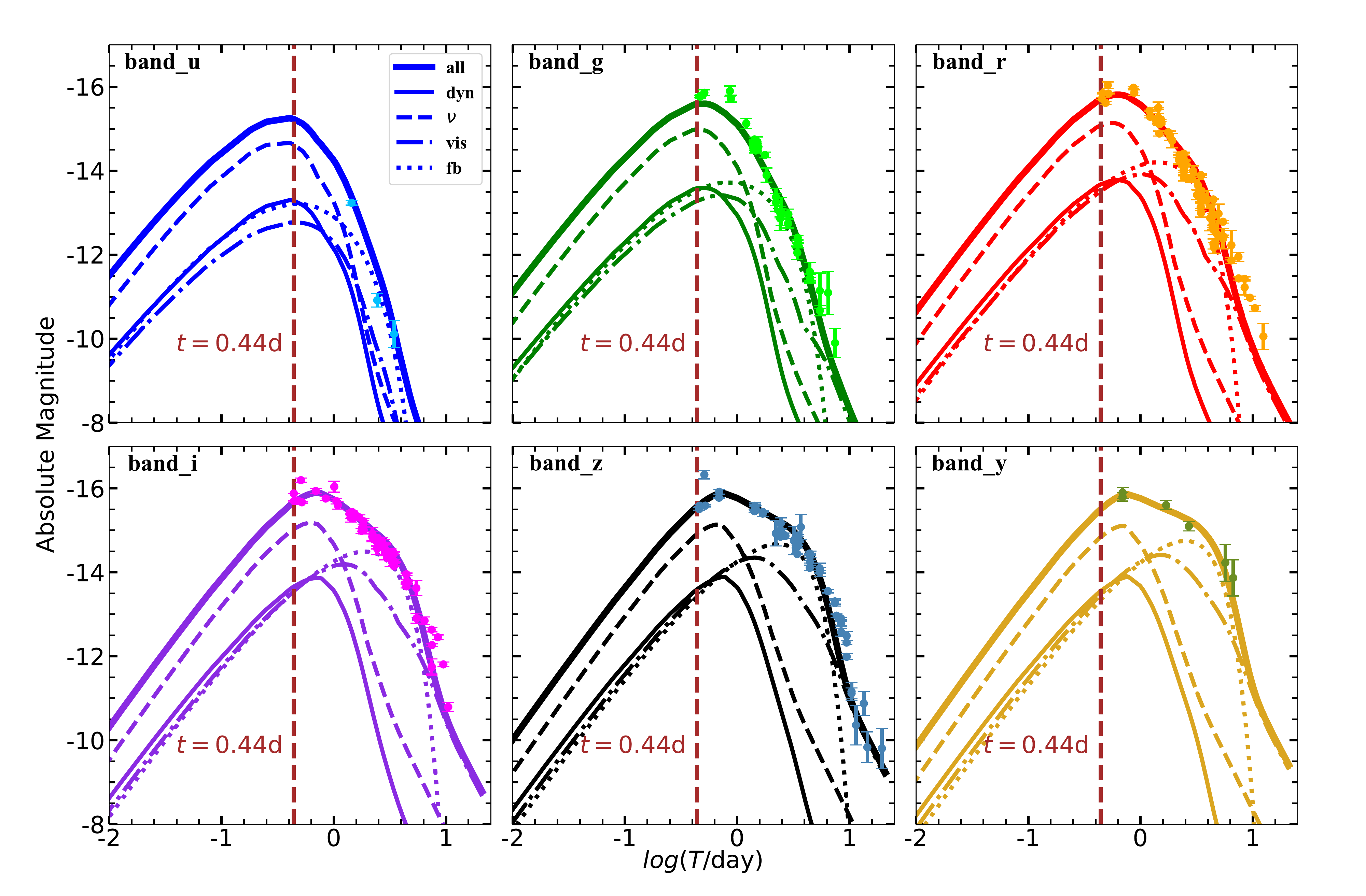}
\caption{
Contributions from different components to the multiband GW170817 LCs in the best-fitting model.  The blue thick solid lines show the predicted LCs at the u, g, r, i, z, and y bands, respectively (as also shown in Fig.~\ref{fig:f1}). In each panel, the thin solid, dash, dot-dashed, and dotted lines represent the contribution from the dynamical ejecta, the neutrino-driven wind, the viscous-driven wind, and the heating from the fallback accretion, respectively; the brown vertical line indicates the time of $0.45$\,days after BNS merged, when the kilonova was first discovered. The data points with errorbars are the multiband observations for AT2017gfo the same as those shown in Fig.~\ref{fig:f1}.
}
\label{fig:f2}
\end{figure*}

Figure~\ref{fig:f2} shows the contribution from different components to the total LCs in the u, g, r, i, z, and y bands, respectively. As seen from this figure, the LC peak appeared at $\sim 0.45-1$\,day after the BNS merger. The exact position of the LC peak depends on which band is considered, and the LC peak for a bluer band emerges earlier than that for a redder band. The neutrino-driven wind contributes the major part of the total luminosity at the peak. The main reason is as follows. The peak luminosity of each component of the kilonova ejecta depends on not only the ejecta mass but also the opacity of the ejecta and velocity, and it can be simply approximated as \citep{Metzger-2019-1-1}
\begin{eqnarray}
L_{\rm peak} \propto \left(\frac{M}{10^{-2}M_\odot}\right)^{0.35} \left(\frac{v}{0.1c}\right)^{\rm 0.65}\left(\frac{\kappa}{1 {\rm cm^2 g^{\rm -1}}}\right)^{\rm -0.65}.
\label{eq:PeakLum_Kne}
\end{eqnarray}
The opacity for the neutrino-driven wind is about $\sim20$ times smaller than that of the viscous-driven wind though the mass of the former one is $\sim20$ times smaller than the latter one, and the velocity of the neutrino-driven wind is also relatively high comparing with that of the latter one. Therefore, the peak luminosity of the neutrino-driven wind is larger than those of the viscous-driven wind by a factor about $\gtrsim2.5$, according to Equation~\eqref{eq:PeakLum_Kne}. The mass of the neutrino-driven wind is slightly smaller than the dynamical ejecta mass, but the opacity for the purple component is substantially smaller than the red component of the dynamical ejecta, which leads to a significant higher peak luminosity from the neutrino-driven wind than that from the dynamical component. At later time, $t\sim 3-10$\,days, the contributions from  the neutrino-driven wind and the dynamical ejecta decline to be negligible and the contributions from the viscous-driven wind and the heating by the fallback accretion begin to dominate the radiation, and the dominance is more evident at redder bands (i-, z-, and y-bands; see bottom panels in Fig.~\ref{fig:f2}). For the red bands, i.e., i-, z-, or y-bands, even a small ``bump'' signature around $t\sim 3-10$\,days appears in the LC because of the contribution from the fallback heating.

\begin{table*}
\centering
\begin{threeparttable}
%
%
\caption{The best-fitting values of model parameters obtained from the fitting to the LCs of GW170817. }
\begin{tabular}{|c|c|c|c|c|c|c|c|c|c|} \hline
$\log\left(\frac{m_{\rm disc}}{M_{\odot}}\right)$ & $\log{\left(\frac{m^{\rm d}_{\rm ej}}{10^{-3}M_{\odot}}\right)}$ & $\theta_{\rm db} (^\circ)$ & $\frac{v_{\rm dyn}}{\textrm{c}}$ & $\frac{t_{\rm n}}{\textrm{day}}$ & $\epsilon_{\rm dec}$ \\ \hline
$-0.96^{+0.01}_{-0.01}$ & $-2.42^{+0.07}_{-0.08}$ & ${30.77^{+3.18}_{-3.31}}$ & $0.285^{+0.037}_{-0.036}$ & $1.09^{+0.35}_{-0.44}$ & $4.33^{+1.48}_{-2.09}$ \\ \hline
$\epsilon_{\rm n}$ & $\log{\left(\frac{\epsilon_{\rm 0}}{{\rm erg~s^{-1}}}\right)}$ & $\log{\left(\frac{\dot{M}_{\rm fb}}{{ M_{\odot}{\rm s}^{-1}}}\right)}$ & $\log{\left(\frac{m_{\rm ph}^{\rm fb}}{10^{-3}M_{\odot}}\right)}$ & $\xi_{\nu}$ & $\xi_{\rm vis}$ \\ \hline
$1.49^{+0.61}_{-0.41}$ & $19.19^{+0.08}_{-0.09}$ & $-2.66^{+0.05}_{-0.05}$ & $-2.54^{+0.23}_{-0.15}$ & $0.014^{+0.005}_{-0.004}$ & $0.260^{+0.050}_{-0.040}$ \\ \hline
$\theta^{\rm \nu b} (^\circ)$ & $\theta^{\rm \nu p} (^\circ)$ & $\theta_{\rm v}(^\circ)$ & $\frac{\kappa_{\rm b}}{\rm cm^2 g^{-1}}$ & $\frac{\kappa_{\rm p}}{\rm cm^2 g^{-1}}$ & $\frac{\kappa_{\rm r}}{\rm cm^2 g^{-1}}$ \\ \hline
$65.3^{+14.3}_{-16.4}$ & $78.2^{+7.9}_{-10.8}$ & $19.02^{+1.60}_{-1.60}$ & $2.53^{+0.40}_{-0.40}$ & $3.90^{+0.86}_{-0.80}$ & $58.96^{+14.50}_{-12.97}$ \\ \hline
$\frac{\kappa_{\rm fb}}{\rm cm^2 g^{-1}}$ & $\frac{T_{\rm f,b}}{\textrm{K}}$ & $\frac{T_{\rm f,p}}{\textrm{K}}$ & $\frac{T_{\rm f,r}}{\textrm{K}}$ & $\frac{T_{\rm f,fb}}{\textrm{K}}$ & $\frac{v_{\rm fb}}{\textrm{c}}$ \\ \hline
$49.58^{+19.45}_{-20.45}$ & $6376^{+263}_{-221}$ & $3147^{+70}_{-68}$ & $430^{+186}_{-162}$ & $1066^{+60}_{-83}$ & $0.130^{+0.011}_{-0.010}$ \\ \hline
$\frac{v_{\nu \rm b}}{\textrm{c}}$ & $\frac{v_{\nu \rm p}}{\textrm{c}}$ & $\frac{v_{\rm visp}}{\textrm{c}}$ & $\frac{v_{\rm visr}}{\textrm{c}}$ & $\sigma$ & \\ \hline
$0.294^{+0.042}_{-0.026}$ & $0.172^{+0.055}_{-0.046}$ & $0.265^{+0.024}_{-0.026}$ & $0.245^{+0.023}_{-0.020}$ & $0.161^{+0.006}_{-0.005}$ & \\ \hline
\end{tabular}
%
%
\begin{tablenotes} 
\item Notes: The quantities from left to right and from top to bottom are the disc mass ($m_{\rm disc}$), the dynamical ejecta mass ($m^{\rm d}_{\rm ej}$), the opening angle of the blue component of the dynamical ejecta ($\theta_{\rm db}$), the velocity of the dynamical ejecta ($v_{\rm dyn}$), the decay time of free neutrons in the outer layer ($t_{\rm n}$), the coefficients of the r-process heating power ($\epsilon_{\rm dec}$, $\epsilon_{\rm n}$ which are fixed by fitting AT2017gfo LCs), the coefficients of the fallback accretion ($\epsilon_0$ and $\dot{M}_{\rm fb}$), the mass of the layer in which the fallback accretion energy passes through ($m^{\rm fb}_{\rm ph}$), the fractions of disc mass ejected in the form of neutrino-driven wind and viscous-driven wind ($\xi_{\nu}$ and $\xi_{\rm vis}$), the opening angle of the blue and purple component of the neutrino-driven wind ($\theta^{\nu \rm b}$ and $\theta^{\nu \rm p}$), the viewing angle of the observer ($\theta_{\rm v}$), the opacities of the blue, purple, red kilonova, and fallback components ($\kappa_{\rm b}$, $\kappa_{\rm p}$, and $\kappa_{\rm b}$, and $\kappa_{\rm fb}$), the temperature floor of blue, purple, and red kilonova, and fallback components ($T_{\rm f,b}$, $T_{\rm f,p}$, $T_{\rm f,r}$, and $T_{\rm f,fb}$), the expanding velocity of the photosphere heated by fallback accretion ($v_{\rm fb}$), the neutrino-driven wind velocities of blue and purple components ($v_{\nu \rm b}$ and $v_{\nu \rm p}$), the viscous wind velocities of purple and red components ($v_{\rm visp}$ and $v_{\rm visr}$), an additional deviation that is added to all measurement errors encompassing additional source of ``white noise".
%
\end{tablenotes}
\label{tab:t1}
\end{threeparttable}
\end{table*}

We note here that \citet{Metzger202133} and \citet{Ishizaki2021185185} showed the fallback from the disc wind can explain the X-ray excess of GW170817 while avoiding overproducing the kilonova emission. Considering the fallback accretion, our model can explain the ``bump'' of the LCs at $t\sim 3-10$\,day well. This again supports the implementation of the fallback accretion into the kilonova modelling.

To close this section, we also note here that the possibility of a magnetized NS remnant of GW170817 is not considered in the above kilonova model. \cite{Yu2018114114} invoked a long-lived remnant NS model for GW170817 and showed that the emission from the remnant NS wind, if it has strong global magnetic field, may be responsible for the later radiation of GW170817. The radiative hydrodynamic simulations by \citet{Kawaguchi2022220213149} showed that the NS remnant with a weak magnetic field may be better consistent with the LCs of GW170817. \citet{Kawaguchi2022220213149} argued that the long-lived massive NS remnants with amplified magnetic fields may be not the majority outcomes of BNS mergers, if the solar element abundance pattern is universal and the BNS mergers are the main sites for the r process. Although the heating from the long-lived remnant NS wind may be considered for a small fraction of the kilonovae, it should not significantly affect our results.

\section{kilonovae from general BNS mergers}
\label{sec:gkilonova}

The LCs for the kilonova associated with a general BNS merger can be also estimated once all the physical parameters for those different components described above are known. Numerical simulations do provide some estimates about these quantities, but not all, thus it is not straightforward to accurately predict the detailed LCs for kilonovae associated with general BNS mergers. One semi-analytical way to predict such LCs may be to combine the numerical simulation results on some parameters and those rest that may be inferred from the observations of GW170817/AT2017gfo. In this section, we first introduce the predictions on dynamical ejecta, wind, and the disc for general BNS mergers by numerical simulations, and then combining with the constraints from GW170817/AT2017gfo observation to predict the LCs for general kilonovae and also the one associated with GW190425. 

\subsection{Dynamical ejecta, wind, and disc predicted for general BNS mergers by NR simulations} 
\label{subsec:NRfittings}

\begin{table} 
\centering 
\caption{The radius ($R_{1.4M_{\odot}}$) of a spherical neutron star with mass $M=1.4M_{\odot}$ but several different EOSs. All of the listed EOSs are consistent with the observation of GW170817 \citep{Bauswein2013131101131101,Coughlin-2019-91-96}.
}
\begin{tabular}{lcccr} \hline
EOS & $R_{\rm 1.4M_{\odot}}/\textrm{km}$ & $R_{\rm 1.6M_{\odot}}/\textrm{km}$ & $M_{\rm TOV}/M_{\rm \odot}$\\ \hline
DD2 & 13.24 & 13.26 & 2.42\\
LS220 & 12.63 & 12.43 & 2.04\\ 
SLy & 11.75 & 11.59 & 2.06\\	\hline
\end{tabular}
\label{tab:t2}
\end{table}

To estimate the multiband LCs of a kilonova, one needs to first estimate the dynamical ejecta, winds, and additional heating from the accretion of fallback material. How much amount of material can be ejected out and how much amount of material resides in the disc are determined by the physical parameters (total mass, mass ratio, etc.) of the BNS merger system with any given EOS. NR simulations for a number of BNS mergers with different EOS \citep[e.g.,][]{Dietrich-2017-105014-105014, Radice-2018-130-130, JuergenKrueger-2020-2002-7728} have shown that these quantities can be roughly given by empirical fitting formulas. Below we briefly introduce these fitting formulas. We assume the masses of the two BNS components are $m_1$ (primary) and $m_2$ (secondary) and the mass ratio $q=m_1/m_2$ ($\geq 1$). The total tidal deformabilities $\widetilde{\Lambda}$ of the BNS system can be given by the following empirical relation \citep[see][]{De-2018-91102-91102}
\begin{equation}
\label{eq:tidaldeform}
\widetilde{\Lambda}\simeq\frac{16a}{13}{\left(\frac{R_{1.4M_\odot}c^2}{G\mathcal{M}}\right)}^{6}f(q), 
\end{equation}
where $a\simeq0.0098$, $R_{1.4M_\odot}$ is the radius of a BNS with mass of $1.4M_{\rm \odot}$, $\mathcal{M}=m_{\rm tot}\left[q/ \left(1+q\right)^2\right]^\frac{3}{5}$ is the chirp mass, $M=m_1+m_2$, and
\begin{equation}
\label{eq:massratio_coeff}
f(q)=q^{\frac{8}{5}}(12-11q+12q^2)(1+q)^{-\frac{26}{5}}.
\end{equation}

The tidal deformability of an NS is sensitive to its radius $R_{1.4M_\odot}$ and thus the EOS. 
Table~\ref{tab:t2} lists the value of $R_{1.4M_{\rm \odot}}$ for NSs with mass of $1.4M_\odot$ but several different EOSs \citep[e.g.,][]{Bauswein2013131101131101, Coughlin-2019-91-96}.
We note that the EOS of NSs is constrained quite tightly by a number of observations, though different works may give somewhat different results. For example, \citet{Pang202222058513} found $R_{1.4M_\odot} =11.98^{+0.35}_{-0.40}$\,km by combining multimessenger observations of GW\,170817 and EM observations of other isolated NSs, which tends to suggest a soft EOS with the maximum mass of non-rotating NSs $ M_{\rm TOV} \sim 2.1M_\odot$ \citep[see also][]{2020PhRvD.101f3007E, 2020NatAs...4..625C, 2021ApJ...918L..28M}. One may directly adopt these constraints rather than a specific EOS to study the properties of kilonova emergent from mergers of BNSs. For our semi-analytical analysis to predict the kilonova properties for the cosmic population of BNS mergers, we consider the SLy and LS220 EOS, which are consistent with the observational constraints obtained by \citet{Pang202222058513} and others (see Table 2).  However, recently \citet{2022ApJ...934L..17R} found that the mass of PSR J0952-0607 is about $2.35\pm 0.17M_\odot$, which suggests $M_{\rm TOV} >2.19M_\odot$, substantially heavier than the constraint given by previous studies. Therefore, there might be still space for a stiff EOS, which gives substantially higher $M_{\rm TOV}$ \citep[see also][]{2020MNRAS.499L..82M} than the constraint of $\sim 2.1M_\odot$ by many studies. For this reason, we also adopt the DD2 EOS, a much stiffer EOS comparing with SLy and LS220, which could represent an extreme case for the EOS of NSs. 

According to the definition of tidal deformability, it is easy to estimate it for individual stars with given mass ratio by using the approximation in \citet{Yagi-2017-1-72}, i.e.,
\begin{equation}
\Lambda_1=q^6\Lambda_2.
\end{equation}
and
\begin{equation}
\Lambda_2=\frac{13}{16}\frac{\left(1+q\right)^5}{\left(12q^4 +q^5 +q^6+12q^7\right)}  \widetilde{\Lambda}.
\label{eq:Tidal}
\end{equation}

For any given compactness, the dynamical ejecta mass can be estimated by adopting the fitting formulas from \citet{JuergenKrueger-2020-2002-7728} and \citet{Coughlin-2019-91-96},
\begin{equation} 
\label{eq:dyn_jkf}
\frac{m^{\rm JKF}_{\rm dyn}}{10^{-3}M_\odot}=\left(\frac{a_1}{C_1}+ a_2\frac{m_2^{n_1}}{m_1^{n_1}}+a_3 C_1\right)m_1+ \left(1\leftrightarrow2\right),
\end{equation}
where
$C_{1,2}=0.360-0.0355\ln{\left(\Lambda_{1,2}\right)}+0.000705\ln{\left(\Lambda_{1,2}\right)^2}$,
and
$a_1=-9.3335$, $a_2=114.17$, $a_3=-337.56$, and $n_1=1.5465$.
\begin{eqnarray}
\label{eq:dyn_cou}
\log\left({\frac{m^{\rm COU}_{\rm dyn}}{M_{\rm \odot}}}\right)=\left[b_1m_1\frac{(1-C_1)}{C_1}+ b_2m_2\left(\frac{m_1}{m_2}\right)^{n_2}+\frac{b_3}{2}\right] \nonumber \\
+ \left(1\leftrightarrow2\right),
\end{eqnarray}
where $b_1=-0.0719$, $b=0.2116$, $b_3=-2.42$, and $n_2=-2.905$. Both equations~\eqref{eq:dyn_jkf} and \eqref{eq:dyn_cou} infer higher dynamical ejecta mass for BNSs with higher mass ratio $(q>1.2)$, especially for those cases with high mass $(m_{\rm tot}>3.0M_{\rm \odot}$ for Equation \ref{eq:dyn_jkf} and $m_{\rm tot}>3.3M_{\rm \odot}$ for Equation \ref{eq:dyn_cou}). However, both these two fitting formulas have their advantages and disadvantages. Equation~\eqref{eq:dyn_jkf} may reflect more rich physics about how the dynamic ejecta mass varying as a function of the BNS mass. At the same mass ratio, less material can be released in the form of the dynamical ejecta if the BNS mass is small $(m_{\rm tot}<2.7M_{\rm \odot})$; while there is less dynamical ejecta for massive BNSs $(m_{\rm tot}>3.2M_{\rm \odot})$, due to its stronger gravitational field binding. However, Equation~\eqref{eq:dyn_jkf} is not sensitive to the BNS mass at the high-mass end and it always infers zero or even negative mass dynamical ejecta when $m_{\rm tot}>3.5M_{\rm \odot}$ and $q<1.3$ for the DD2 EOS ($m_{\rm tot}>3.2M_{\rm \odot}$ and $q<1.3$ for the SLy EOS). Equation~\eqref{eq:dyn_cou} takes a logarithmic form, which can avoid inferring zero or negative mass for the dynamical ejecta. However, Equation~\eqref{eq:dyn_cou} indicates the dynamical ejecta mass $m^{\rm COU}_{\rm dyn}$ to monotonically increase with BNS mass, neglecting the strong gravitation field binding at the high BNS mass end. 
Therefore, we take the mean value of $m^{\rm JKF}_{\rm dyn}$ and $m^{\rm COU}_{\rm dyn}$ as the inferred dynamical ejecta mass, i.e.,
\begin{equation} 
\label{eq:mej_dyn}
m^{\rm d}_{\rm ej}=wm^{\rm JKF}_{\rm dyn}+(1-w) m^{\rm COU}_{\rm dyn}.
\end{equation}
where we adopt the weight $w = (C_1-C_{\rm min})/(C_{\rm max}-C_{\rm min})$. The quantity $C_{\rm max}$ is the maximum compactness of all the BNS systems given a specific EOS and the quantity $C_{\rm min} \simeq 0.1$ is the minimum compactness. When $C_{1} \rightarrow C_{\rm max}$ (or $w \rightarrow 1$),  $m^{\rm JKF}_{\rm dyn}$ term dominates, whereas $m^{\rm COU}_{\rm dyn}$ becomes the leading term when $C_{1} \rightarrow C_{\rm min}$. Therefore, the dynamical ejecta mass $m^{\rm d}_{\rm ej}$ decreases rapidly with increasing total mass of BNS at the high-mass region ($m_{\rm tot}>3.5M_{\rm \odot}$) and it is $\sim m^{\rm JKF}_{\rm dyn}$ and close to $0$, while it increases with the BNS total mass at the low-mass region ($m_{\rm tot}<3.5M_{\rm \odot}$).
We consider a scatter of $\sigma_{\rm dyn}=0.004M_\odot$ to the above $m^{\rm d}_{\rm ej}$ estimate and randomly assign a deviation to the $m^{\rm d}_{\rm ej}$ estimate according to a Gaussian distribution $N(m^{\rm d}_{\rm ej},0.004M_\odot)$ \citep[cf.][]{Dietrich-2017-105014-105014, Coughlin-2019-91-96,JuergenKrueger-2020-2002-7728}. Note that when $m^{\rm JKF}_{\rm dyn}$ in Equation~\eqref{eq:dyn_jkf} is negative, the value of $m^{\rm d}_{\rm ej}$ could  be  also negative. We reject negative $m^{\rm d}_{\rm ej}$ inferred from the above process and reassign $m^{\rm d}_{\rm ej}$  according to the scatter distribution until a positive $m^{\rm d}_{\rm ej}$ is obtained. The distribution of the dynamical ejecta mass at the high $m_{\rm tot}$ region obtained in this way is $\sim10^{-3}-10^{-2}M_{\rm \odot}$, consistent with that obtained by simulations  \citep{Dietrich-2017-105014-105014, JuergenKrueger-2020-2002-7728}. 

As for the disc mass, the fitting formulas based on NR simulations have also been provided by \citet[][equation~\eqref{eq:disc_rea} below]{Coughlin-2019-91-96} and \citet[][equation~\eqref{eq:disc_cf} below]{JuergenKrueger-2020-2002-7728}, respectively. In  \cite{Coughlin-2019-91-96}, the disc mass is linked to the threshold mass $m_{\rm thr}$ above which the remnant directly collapses into a BH \citep{Bauswein2013131101131101}, i.e.,
\begin{equation} 
\label{eq:Mthr}
M_{\rm thr}=kM_{\rm TOV},
\end{equation}
and
\begin{equation} 
\label{eq:compact16}
k=j_1C^{\star}_{1.6}+j_2, C^{\star}_{1.6}=\frac{GM_{\rm TOV}}{c^2R_{1.6M_{\rm \odot}}}
\end{equation}
where $j_1=-3.606$ and $j_2=2.380$. Note that $R_{1.6M_{\rm \odot}}$, the radius of the NS with mass $1.6M_{\odot}$, and the corresponding $C^{\star}_{1.6}$ is the compactness of the NS with mass $1.6M_{\odot}$ is adopted in the above equation, for consistency with that in \cite{Bauswein2013131101131101}. With the threshold mass $M_{\rm thr}$ for BNSs to collapse into BHs, we may obtain the disc mass according to \citet{Coughlin-2019-91-96}, i.e.,
\begin{equation} 
\label{eq:disc_rea}
\log\left(\frac{m_{\rm disc}^{\rm COU}}{M_\odot}\right) = \max \left[ -3, d_1\left(1+d_2\tanh{\left[\frac{d_3-m_{\rm tot}/M_{\rm thr}}{d_4}\right]}\right) \right],
\end{equation}
where
$d_1=-31.335$, $d_2=-0.976$, $d_3=1.0474$, $d_4=0.05957$; or alternatively according to \citet{JuergenKrueger-2020-2002-7728}
\begin{equation}
\label{eq:disc_cf}
\frac{m_{\rm disc}^{\rm JKF}}{M_{\rm \odot}}=m_1\left[\max{\left( e_1 C_1+ e_2,\
5\times 10^{-4}\right)}\right]^{n_3},
\end{equation}
where
$e_1=-8.1324$, $e_2=1.4820$, and $n_3=1.7784$.

We note that, on the one hand, the fitting formula given by \citet{JuergenKrueger-2020-2002-7728} (equation~\ref{eq:disc_cf}) cannot apply to the cases with low compactness $C_1<0.135$, while Equation~\eqref{eq:disc_rea} given by \citet{Coughlin-2019-91-96} approaches asymptotically to a physically reasonable value of disc mass ${\sim}0.2M_{\odot}$ for such cases. On the other hand, Equation~\eqref{eq:disc_rea} predicts nearly no accretion disc formed for cases with low deformability $\widetilde{\Lambda}<250$, which is in contradiction with the simulation results obtained by \citet{Kiuchi-2019-31-31}, while Equation~\eqref{eq:disc_cf} predicts disc masses consistent with NR simulations. Considering the advantages and disadvantages of these two fitting formulas, we combine them together by introducing a weight parameter $w$ in order to avoid the disadvantages of these two formulas for cases with low compactness or deformability as following
\begin{equation}
\label{eq:disc}
m_{\rm disc}=w m_{\rm disc}^{\rm JKF}+(1-w)m_{\rm disc}^{\rm COU},
\end{equation}
where we also adopt the same weight $w = (C_1-C_{\rm min})/(C_{\rm max}-C_{\rm min})$ as in Equation~\eqref{eq:mej_dyn}. In the case of low compactness, the term $m_{\rm disc}^{\rm COU}$ dominates, while the term $m_{\rm disc}^{\rm JKF}$ becomes dominant in other cases. When $C_1$ is close to $C_{\rm max}\simeq0.22-0.25$ for any given EOS, there is no disc formed according to Equation~\eqref{eq:disc_cf} \citep{JuergenKrueger-2020-2002-7728}. We also assume a deviation for the disc mass as follows \citep{Radice-2018-130-130, Coughlin-2019-91-96, JuergenKrueger-2020-2002-7728}
\begin{equation}
\label{eq:disc_residual}
\Delta m_{\rm disc}=0.5m_{\rm disc}+5\times10^{-4}.
\end{equation}
The treatment of the deviation of the disc mass is similar as that of dynamical ejecta mass. We generate a Gaussian distribution $N(m_{\rm disc},\Delta m_{\rm disc})$ and randomly sample a value as the disc mass from this distribution but reject negative $m_{\rm disc}$ generated by this process.

We refer to \citet{Dietrich-2017-105014-105014} for the opening angle of dynamical ejecta and \citet{Coughlin-2019-91-96} for its total velocity, respectively, i.e., 
\begin{equation}
\label{eq:theta_dyn}
\theta_{\rm db} \simeq \frac{-2^\frac{4}{3} v_\rho^2 +2^\frac{2}{3} \left[ v_\rho^2 \left(3v_z+ \sqrt{9v_z^2 +4v_\rho^2}\right)\right]^\frac{2}{3}}{\left[ v_\rho^5 \left(3v_z+\sqrt{9v_z^2+4v_\rho^2}\right)\right]^\frac{1}{3}},
\end{equation} 
\begin{equation}
\label{eq:v_dyn}
v_{\rm dyn}^{\rm fit} \simeq \left[f_1\left(1+f_3 C_1\right) \frac{m_1}{m_2}+\frac{f_2}{2}\right]+\left(1\leftrightarrow2\right)
\end{equation} 
where $f_1=-0.3090$, $f_2=0.657$, $f_3=-1.879$, $v_\rho$ and $v_z$ are the radial and perpendicular components of dynamical ejecta velocity given by 
\begin{equation}
v_\rho=\left[g_1\left(\frac{m_1}{m_2}\right)\left(1+g_3 C_1\right)\right]+\left(1\leftrightarrow2\right)+g_2,
\label{eq:v_dyn_pho}
\end{equation} 
and
\begin{equation}
\label{eq:v_dyn_z}
v_z=\left[h_1 \left(\frac{m_1}{m_2}\right)\left(1+ h_3 C_1\right)\right]+\left(1\leftrightarrow2\right)+h_2,
\end{equation} 
respectively, where
$g_1=-0.219479$, $g_2=0.444836$,  $g_3=-2.67385$,
$h_1=-0.315585$, $h_2=0.63808$, and $h_3=-1.00757$. 
We adopt Equations~\eqref{eq:v_dyn}-\eqref{eq:v_dyn_z} to obtain the opening angle of the blue dynamical ejecta.

\subsection{Settings for other kilonova parameters}
\label{subsec:parasettings}

To generate the LCs for a general kilonova, we also need to set those model parameters, which may be not directly given by current NR simulations as $m_{\rm ej}^{\rm d}$, $m_{\rm disc}$, $v_{\rm d}$, and $\theta_{\rm db}$ described in Section~\ref{subsec:NRfittings} (see also Section~\ref{subsec:LCs}). Below we describe the settings for each of those model parameters.

For the fallback process, the total fallback mass is directly related to the ejecta mass and velocity and the total mass of the BNS merger considering the effects of gravitational binding and escape as pointed out in \citet[][]{Ishizaki-2021-13-13}, i.e., 
\begin{eqnarray}
{M}_{\rm fb,tot} & = & 5.72\times10^{\rm -3}M_{\rm \odot}\left(\frac{t_{\rm 0}}{5\rm s}\right)^{\rm -0.2}\left(\frac{m_{\rm tot}}{2.7M_{\rm \odot}}\right)^{\rm 0.2} \nonumber \\
& & \left(\frac{m_{\rm ej}}{0.08M_{\rm \odot}}\right)\left(\frac{v_{\rm ej}}{0.04c}\right)^{\rm -0.6}, 
\label{eq:fbmass}
\end{eqnarray}
where $m_{\rm tot}=m_1+m_2$ representing the total mass of the BNS merger, $m_{\rm ej}$ and $v_{\rm ej}$ representing the ejecta mass and velocity, and $t_0$ representing the starting time for the fallback accretion after the merger. The total fallback mass from the dynamical ejecta and the disc wind should follow the same relation as Equation~\eqref{eq:fbmass} though with different normalization because the mass density profile of the disc wind is similar to that of the dynamical ejecta \citep{Ishizaki2021185185}. However, the effective time-scale for the fallback accretion due to the disc wind is much longer than that due to the dynamical ejecta, and the disc fallback effects only evident in the X-ray band after $100$\,days. Therefore, we only consider the fallback of the dynamical ejecta in the present paper as we focus on the optical-infrared emission from kilonovae at their early evolution time (a few ten days), and adopt Equation~\eqref{eq:fbmass} to estimate $M_{\rm fb,tot}$ for the dynamical ejecta by using $m_{\rm ej}=m^{\rm d}_{\rm ej}$ and $v_{\rm ej} = v_{\rm dyn}^{\rm fit}$ given in Section~\ref{subsec:NRfittings}. With this $M_{\rm fb,tot}$, one can infer $\dot{M}_{\rm fb}(0.1{\rm s})$ in Equation~\eqref{eq:fb} if $t_0$ is given. To avoid of the uncertainty in the setting of $t_0$, we introduce simple linear scaling relations to obtain the fallback accretion rate and the mass for the photosphere due to the heating from fallback accretion (presuming from the dynamical ejecta) by using the fitting result on $\dot{M}_{\rm fb}$ for GW170817 (see Table~\ref{tab:t1}), i.e., 
\begin{equation}
\dot{M}_{\rm fb}(0.1{\rm s}) = \dot{M}^{\rm 170817}_{\rm fb,obsf} \left(\frac{M_{\rm fb,NR}}{M_{\rm fb,NR}^{\rm 170817}}\right),
\label{eq:accretion rate}
\end{equation}
and
\begin{equation}
\begin{aligned}
m_{\rm ph}^{\rm fb} = m_{\rm ph}^{\rm fb,170817} \left(\frac{M_{\rm fb,NR}}{M_{\rm fb,NR}^{\rm 170817}}\right),
\label{eq:photosphere fallback}
\end{aligned}
\end{equation}
where $M_{\rm fb,NR}$ and $M_{\rm fb,NR}^{\rm 170817}$ are the total fallback mass estimated by adopting Equation~\eqref{eq:fbmass} for a general BNS merger and GW170817, respectively, $\dot{M}^{170817}_{\rm fb,obsf}$ and $m_{\rm ph}^{\rm fb,170817}$ are the best-fitting values of $\dot{M}_{\rm fb}$ and $m^{\rm fb}_{\rm ph}$ for GW170817 (see Table~\ref{tab:t1}), respectively. 

Kilonova emission does depend on the viewing angle $\theta_{\rm v}$, though the dependence is not as significant as the afterglow. Normally $\theta_{\rm v}$ cannot be well determined solely from the GW signal, but the probability distribution of $\theta_{\rm v}$ for a sample of BNS mergers detected by detectors can be described by \citep[see][]{Schutz-2011-125023-125023}
\begin{equation}
P(\theta_{\rm v})=0.076076(1+6\cos^2{\theta_{\rm v}}+\cos^4{\theta_{\rm v}})^{3/2}\sin{\theta_{\rm v}}.
\label{eq:view_angle_distr}
\end{equation}
According to this distribution, the mean value for $\theta_{\rm v}$ is $\left<\theta_{\rm v} \right>\sim 38^\circ$ or $142^\circ$ considering the symmetry, the peak value of $\theta_{\rm v}$ is $\sim 31^\circ$ or $149^\circ$. Hereafter, we either set $\theta_{\rm v}$ as its mean value or some specific values when discussing a few typical cases, or following the above distribution when generating mock samples for the GW detected ones. 

Currently, it is difficult to give detailed and accurate settings for the rest $21$ model parameters (excluding $\sigma$, which is irrelevant for predicting LCs), such as the velocity, opening angle, opacity, and the minimum temperature of the corresponding photosphere for different ejecta components, the fractions of the neutrino- and viscous-driven wind mass to the disc mass, etc., according to current NR simulations. Probably these parameters are also different for different BNS mergers with different physical parameters and orbital configurations, which may be inferred in future by much more comprehensive 
dynamical and radiative NR simulations for a broad range of BNS mergers and thus enable the vastly preferred model grid. 
Nevertheless, we assume these model parameters are the same as those obtained from the best fits to the LCs of GW170817 (see Table~\ref{tab:t1}).

\subsection{Estimations of the GW190425 LCs}
\label{subsec:De_GW190425}

GW190425 is the second BNS merger detected by LIGO/VIRGO. The GW observations of  GW190425 indicate the BNS total mass ${\sim}3.2$-$3.4M_{\odot}$, substantially larger than any known Galactic BNSs \citep{Abbott-2020-3-3}. The uncertainty in the estimation of the GW190425 mass ratio $q$ is considerable, and $1{\leq}q{\leq}1.27$ if adopting the low-spin prior, while $1{\leq}q{\leq}2.5$ if adopting the high-spin prior \citep{Abbott-2020-3-3}. No EM counterpart was detected for GW190425 though many efforts have been made for this BNS merger, which is partly due to the large  uncertainty in the source localization and partly due to the faintness of the EM counterpart as its luminosity distance is much larger than that of GW170817. The observational searches for the kilonova associated with GW190425 suggest that the upper limits for the absolute magnitudes for the kilonova are $\sim-15.0^{-0.5}_{+0.3}$\,mag in the g  and r bands, $\sim-20.5^{-0.4}_{+0.3}$\,mag in the J-band \citep{Coughlin-2019-19-19}, $\sim-13.5^{-1.3}_{+1.0}$\,mag in the i-band \citep{Hosseinzadeh-2019-4-4}, and $-14.9_{-0.8}^{+1.3}$\,mag in the u band \citep{Tohuvavohu201911}, respectively. These upper limits for the different absolute magnitudes are obtained by adopting the luminosity distance of GW190425 ($159^{+69}_{-71}$Mpc) \citep[see][]{Foley-2020-190-198}.

To estimate the LCs of GW190425, we assume the total BNS mass is either $3.2M_{\rm \odot}$ or $3.4M_{\rm \odot}$ and the mass ratio $q$ is either $1$ or $1.27$ according to the GW observations. We also assume either a stiff DD2 EOS or a softer SLy EOS, which both are compatible with the GW observation. The viewing angle is assumed to be $\sim \left< \theta_{\rm v} \right>=38.23^\circ$. The combination of these parameter choices gives eight models for GW190425. For these eight models, the disc mass, the disc wind mass, and the dynamical ejecta mass inferred from Equations~\eqref{eq:disc} and \eqref{eq:mej_dyn} for GW190425 are in the range of $\sim 10^{-3}-10^{-2} M_{\rm \odot}$, $\sim 10^{-4}-10^{-2}M_{\rm \odot}$, and $\sim10^{-4}-10^{-2}M_{\rm \odot}$, respectively, which are compatible with those in previous works \citep{Barbieri-2020-2002-9395, Raaijmakers2021269269}. We calculate the LCs of GW190425 for these eight models according to the processes introduced in Sections~\ref{sec:model}, \ref{sec:gkilonova}, and \ref{subsec:parasettings}, and find they are all compatible with observations, though some models predict slightly brighter kilonova and some others predict slightly fainter kilonova. 

\begin{figure*}
\includegraphics[width=0.8\textwidth]{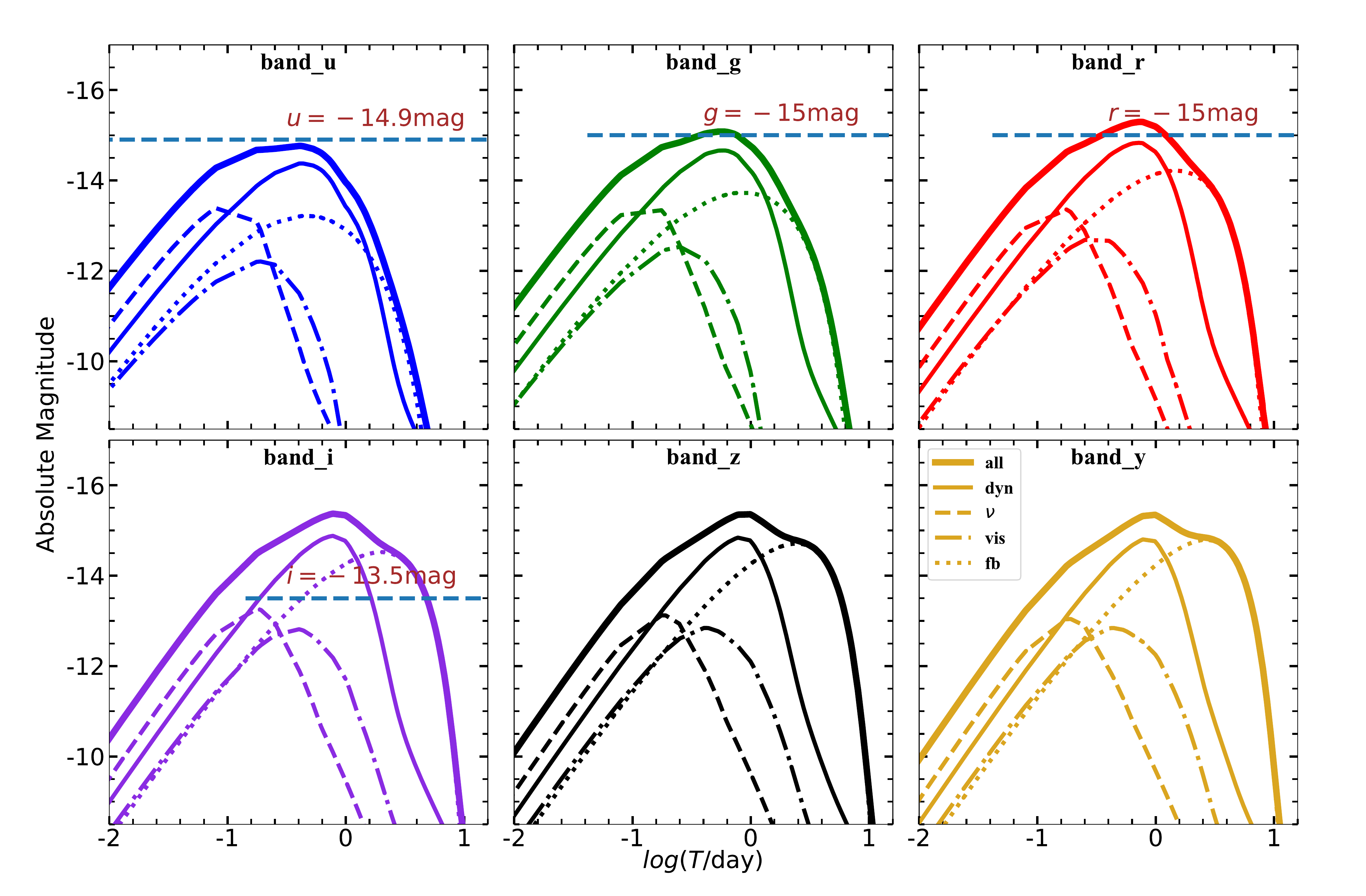}
\caption{
Legends are similar to those for Fig.~\ref{fig:f2}, except that the LCs are generated for BNS mergers with the same properties as GW190425 by assuming the DD2 EOS (see Section.~\ref{subsec:De_GW190425}, choosing $m_{\rm tot}=3.3M_{\rm \odot}$, and $q=1.27$ which is one of the brightest model). EM observations did not find the kilonova associated with GW190425, and the horizontal dashed lines denote the limiting magnitudes of the u, g, r, and i bands, respectively, for those searches conducted for detecting its EM counterparts \citep{Coughlin-2019-19-19, Hosseinzadeh-2019-4-4, Tohuvavohu201911, Foley-2020-190-198}.
}
\label{fig:f3}
\end{figure*}

Figure~\ref{fig:f3} shows eight the LCs for the kilonova associated with GW190425 in the u, g, r, i, z, and y bands, predicted from one of the eight models (with $m_{\rm tot}=3.3M_\odot$, $q=1.27$, DD2 EOS). In this model, we have $m_{\rm disc}\simeq3.2\times10^{-3}M_{\rm \odot}$ and $m^{\rm d}_{\rm ej}\simeq4.2\times10^{-3}M_{\rm \odot}$, and the resulting peak luminosity is one of the highest among those resulting from all the eight models. As seen from Figure~\ref{fig:f3}, the contribution from the fallback engine is significant to the total luminosity around the peaks of the LCs, and even dominant after the peaks; the contribution from the neutrino-driven wind is significant only when the elapsing time is shorter than $\sim 0.1$\,day, substantially before the LCs reach its peaks; the contribution from the dynamical ejecta to the total luminosity is comparable to that from the fallback accretion around the LC peaks. The non-significant contribution from the disc wind in this case, different from that for GW170817 shown in Figure~\ref{fig:f1}, mainly because the remnant disc mass  $(3.2 \times 10^{-3}M_{\rm \odot})$ is $\sim30\%$ of that of GW170817 and thus the luminosity contributed by the disc wind in GW190425 is minor. The relative significance of different ejecta components define the shapes of the LCs at different bands, which may be used to constrain the ejecta properties by future observations of GW170817- and GW190425-like kilonovae.

\subsection{LCs for general kilonovae and its dependence on the physical properties of BNS mergers}
\label{subsec:PhyTempLCs}

The LCs for any BNS merger can be predicted according to the kilonova model described in Sections~\ref{sec:model}, \ref{subsec:NRfittings}, and \ref{subsec:fallback}. Obviously, the kilonova brightness and LC shapes depend on the physical properties of the BNS merger, especially, the EOS, the total mass, and the mass ratio. In this section, we analyse the dependence by producing LCs for a large number of kilonovae with a broad range of BNS physical properties. 

\subsubsection{Example LCs}

For illustration, we first show the LCs for a number of BNS mergers with different physical properties by using the above kilonova model under different EOS assumptions. We choose three different EOSs, i.e., DD2 (stiff), LS220 (relatively soft), SLy (very soft). We assume that the total mass of the merger is either $m_{\rm tot}= 2.7M_{\odot}$ or $3.3M_{\odot}$, nearly equal to that of GW170817 or GW190425; the mass ratio is either $q=1.05$ or $1.25$; the viewing angle $\theta_{\rm v}$ is fixed at $19.02^\circ$, the same as GW170817. Some model parameters to produce these LCs are listed in Table~\ref{tab:t3}.

\begin{figure*}
\includegraphics[width=6.5in]{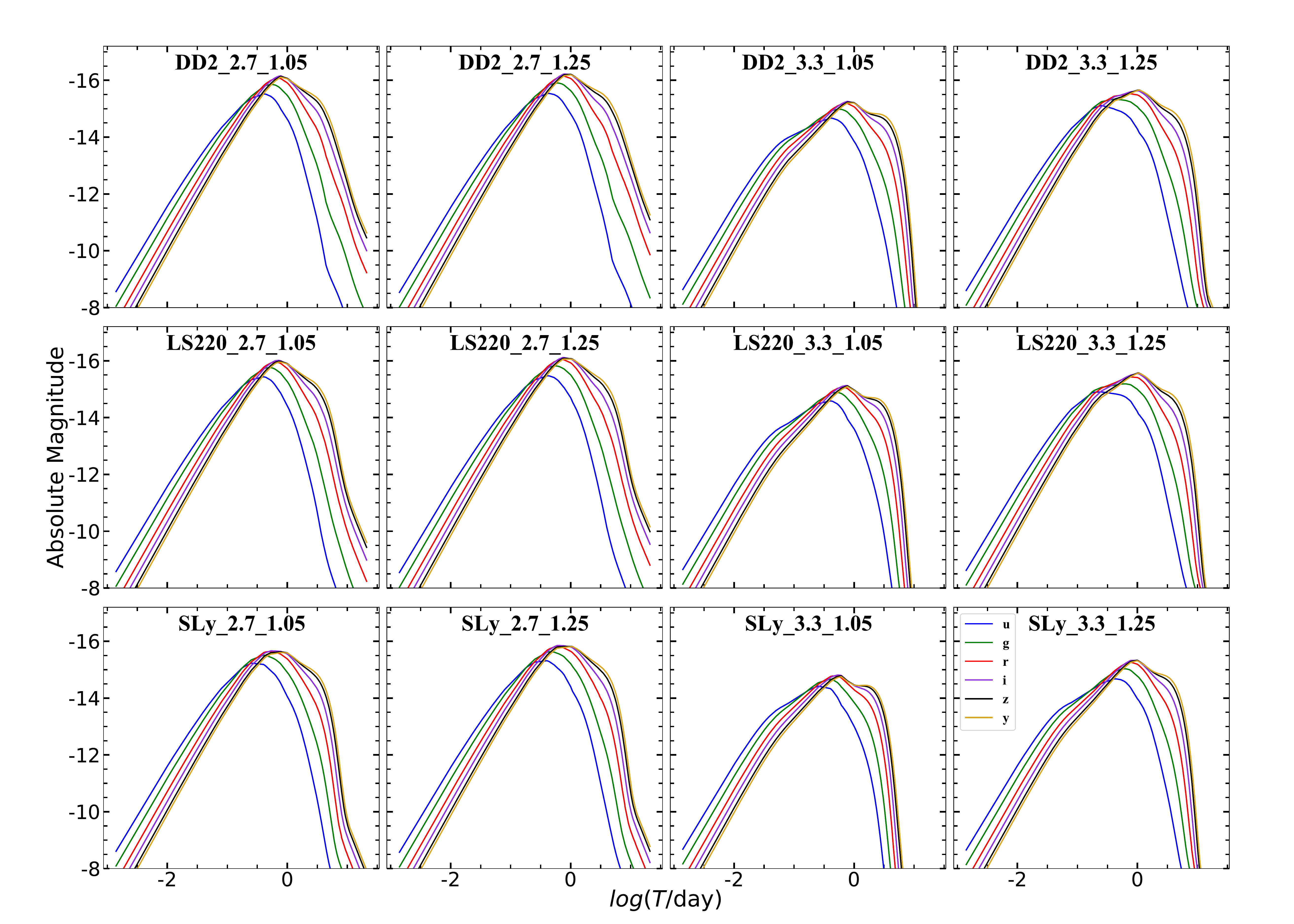}
\caption{Multiband LCs (magnitude versus time) obtained from the model introduced in Section \ref{sec:model} by assuming different binary parameters and EOSs. Here the time $t$ is assumed to be zero when the BNS merges. From left-hand to right-hand column, the total mass and mass ratio of the binaries are set as ($2.7M_\odot, 1.05$), ($2.7M_\odot, 1.25$), ($3.3M_\odot, 1.05$), and ($3.3M_\odot, 1.25$), respectively. From top to bottom columns, the EOSs are assumed to be DD2, LS220, and SLy, respectively. In each panel, the blue, orange, green, red, purple, and brown solid lines represent the LCs at u, g, r, i, z, and y band, respectively. 
}
\label{fig:f4}
\end{figure*}

\begin{table*}
\centering
\caption{
The values adopted for the disc mass, dynamical ejecta mass, dynamical ejecta velocity, opening angle of the blue dynamical ejecta component applied in generating the LCs shown in Fig.~\ref{fig:f4}. All of these quantities are estimated by using the fitting formulas obtained from the NR simulations as described in Section~\ref{subsec:NRfittings}.
}
\begin{center}
\begin{tabular}{c| cccc| cccc| cccc} \hline \hline
EOS & \multicolumn{4}{c}{DD2} & \multicolumn{4}{c}{LS220} & \multicolumn{4}{c}{SLy} \\ \hline 
$\frac{m_{\rm tot}}{M_{\rm \odot}}$ & 2.7 & 2.7 & 3.3 & 3.3 & 2.7 & 2.7 & 3.3 & 3.3 & 2.7 & 2.7 & 3.3 & 3.3 \\ \hline
$q$ & 1.05 & 1.25 & 1.05 & 1.25 & 1.05 & 1.25 & 1.05 & 1.25 & 1.05 & 1.25 & 1.05 & 1.25 \\ \hline
$\frac{m_{\rm disc}}{10^{-2}M_{\rm \odot}}$ & $15.81$ & $18.32$ & $0.07$ & $1.79$ & $9.71$ & $12.67$ & $0.03$ & $0.78$ & $3.23$ & $5.63$ & $0.03$ & $0.04$ \\ \hline
$\frac{m_{\rm dyn}}{10^{-2}M_{\rm \odot}}$ & $0.27$ & $0.41$ & $0.38$ & $0.71$ & $0.35$ & $0.52$ & $0.27$ & $0.69$ & $0.38$ & $0.60$ & $0.12$ & $0.50$ \\ \hline
$\frac{v_{\rm dyn}}{\textrm{c}}$ & 0.22 & 0.21 & 0.25 & 0.25 & 0.22 & 0.22 & 0.26 & 0.26 & 0.24 & 0.23 & 0.28 & 0.27 \\ \hline
$\theta_{\rm dyn}$ & $58.8^{\circ}$ & $54.1^{\circ}$ & $58.6^{\circ}$ & $54.7^{\circ}$ & $58.8^{\circ}$ & $54.3^{\circ}$ & $58.5^{\circ}$ & $54.8^{\circ}$ & $58.7^{\circ}$ & $54.5^{\circ}$ & $58.4^{\circ}$ & $54.9^{\circ}$ \\ \hline \hline
\end{tabular}
\end{center}
\label{tab:t3}
\end{table*}

Figure~\ref{fig:f4} shows the resulting LCs for $12$ BNS mergers by the combination of the EOSs and the physical properties set above. Given the same mass ratio and EOS, the kilonovae from BNS mergers with $m_{\rm tot}=3.3M_\odot$ are fainter (e.g., by $\sim0.5-1.0$ \,mag at the peak) compared with those from mergers with $m_{\rm tot}=2.7M_\odot$ (the first and third columns or the second and fourth columns in Fig.~\ref{fig:f4}), because the high-mass BNS merger tends to promptly collapse to a BH and form a small remnant disc with less significant disc wind and dynamical ejecta. In addition, the kilonova from a BNS merger with $m_{\rm tot} =3.3M_\odot$ declines more rapidly than that from those with $m_{\rm tot} =2.7M_\odot$. The main reason is as follows. The later time emission (after the peak emission) of the former one may first be dominated by the neutrino- and viscous-driven wind components and then by the fallback accretion, while that of the latter one is soon dominated by the fallback accretion. The r-process engine of the disc wind leads to a more gentle decline slope, but the fallback accretion produces a steeply descending tail LC. Therefore, the resulting LCs of the kilonovae from BNS mergers with higher total mass declines more rapidly.
Given the same $m_{\rm tot}$ and EOS, the peak luminosity of the resulting kilonova in the case with $q=1.25$ is higher than that with $q=1.05$ (by a magnitude of $\sim0.5$\,mag) when $m_{\rm tot}$ is large (see the 3rd and 4th columns of Fig.~\ref{fig:f4}), because the resulting disc mass from the merger with $q=1.25$ is larger than that with $q=1.05$ if $m_{\rm tot}$ is large. However, $q$ does not have much effect on the LCs when $m_{\rm tot}$ is small (see the first and second columns in Fig.~\ref{fig:f4}). The reason is the inferred disc mass converges to $\sim 0.2M_{\rm \odot}$ when $m_{\rm tot}$ decreases to the small-mass end according to Equation~\eqref{eq:disc_rea}, which thus lead to the peak luminosities of BNS mergers with different $q$ being restricted almost in the same range. For the BNS mergers with the same $m_{\rm tot}$ and $q$ but different EOSs, those with softer EOSs (e.g., SLy) have less significant ejecta and thus result in fainter kilonovae (with peak magnitude fainter by $\sim0.5$\,mag) comparing with those with stiff EOSs (e.g., DD2).

\begin{figure*}
\includegraphics[width=6.5in]{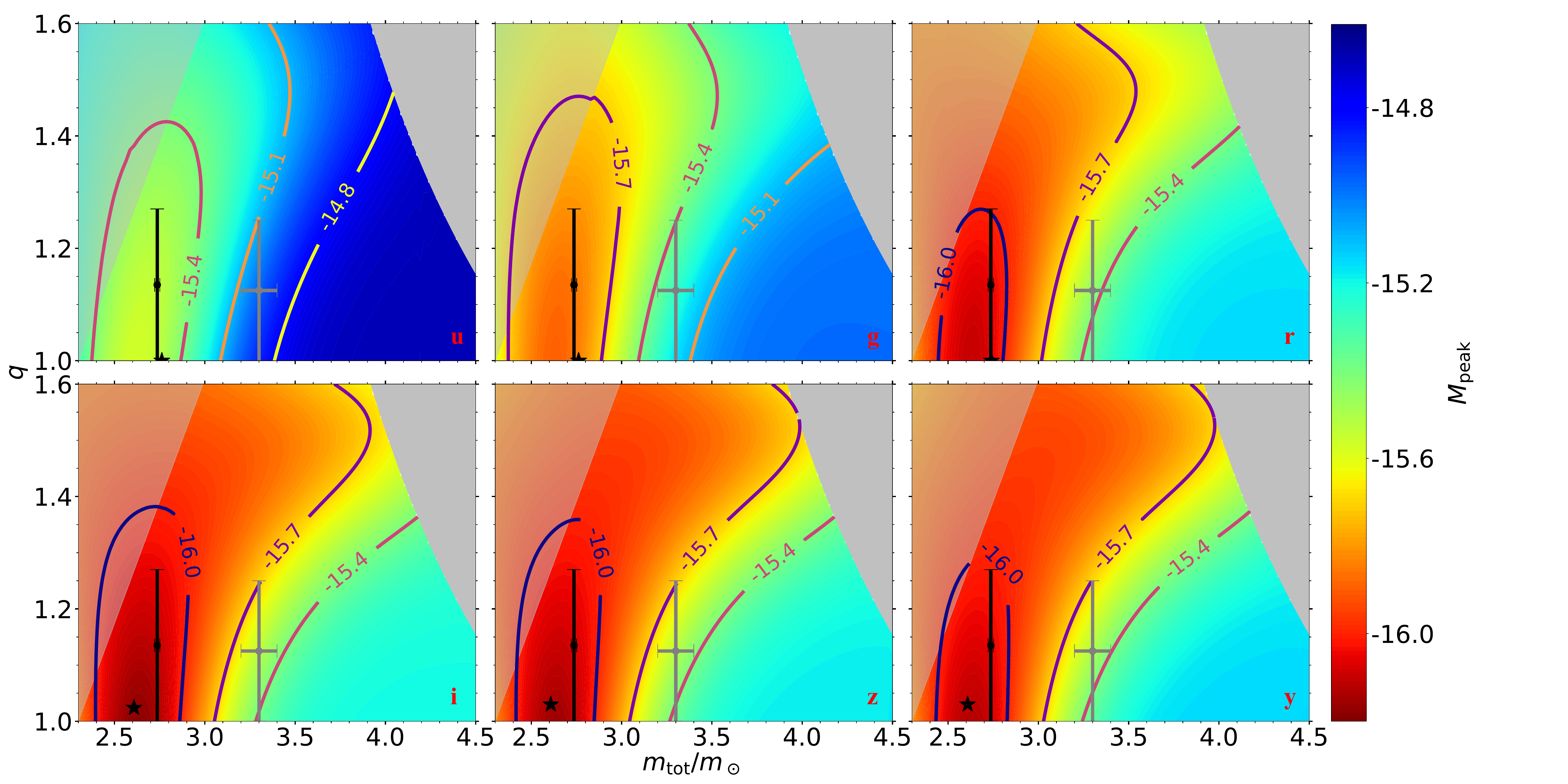}
\caption{Distributions of the multiband absolute magnitudes at the peak luminosity on the $m_{\rm tot}-q$ plane by assuming the DD2 EOS (see Section~\ref{subsec:Mpeak_Mtot_q}). Panels from left to right and from top to bottom show the results for the u, g, r, i, z, and y bands, respectively. The view angle is chosen to be the same as that of GW170817, i.e., $19^\circ$. The colourbar at the right side indicates the value of the absolute magnitude at different bands. In each panel,  the colour contour lines show those kilonovae, predicted from BNS mergers with different ($m_{\rm tot}$, $q$), with the same peak absolute magnitude as marked by the numbers associated with the lines; the grey regions are excluded as they are not allowed by the settings of the minimum and maximum masses for NSs; the black `star' points represent the most luminous kilonovae with the peak luminosities as $-15.57$, $-15.89$, $-16.10$, $-16.18$, $-16.16$, and $-16.13$ from left to right and from top to bottom, respectively; the black and grey symbols with associated errorbars indicate the location of GW170817 (left) and GW190425 (right) on the $m_{\rm tot}-q$ plane, respectively \citep[see][]{Coughlin-2019-91-96, Abbott-2020-3-3}.
}
\label{fig:f5}
\end{figure*}

\begin{figure*}
\includegraphics[width=6.5in]{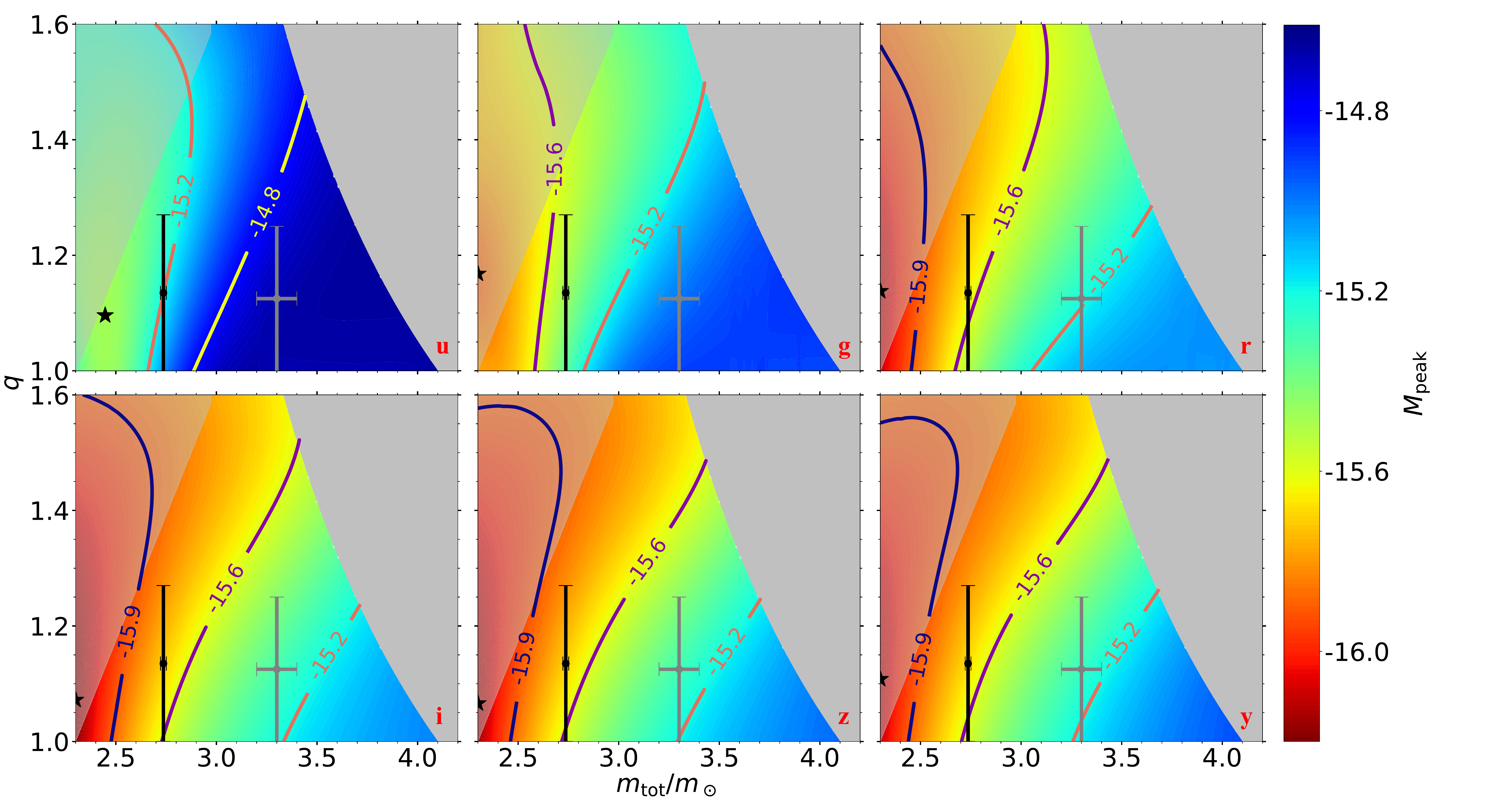}
\caption{
Legends are the same as that for Fig.~\ref{fig:f5}, except that the EOS for neutron stars is assumed to be the SLy EOS, with the kilonovae peak luminosities being $-15.47$, $-15.84$, $-16.15$, $-16.19$, $-16.14$, and $-16.13$ in panels from left to right and from top to bottom, respectively.
}.
\label{fig:f6}
\end{figure*}

\begin{figure*}
\centering
\includegraphics[width=6.5in]{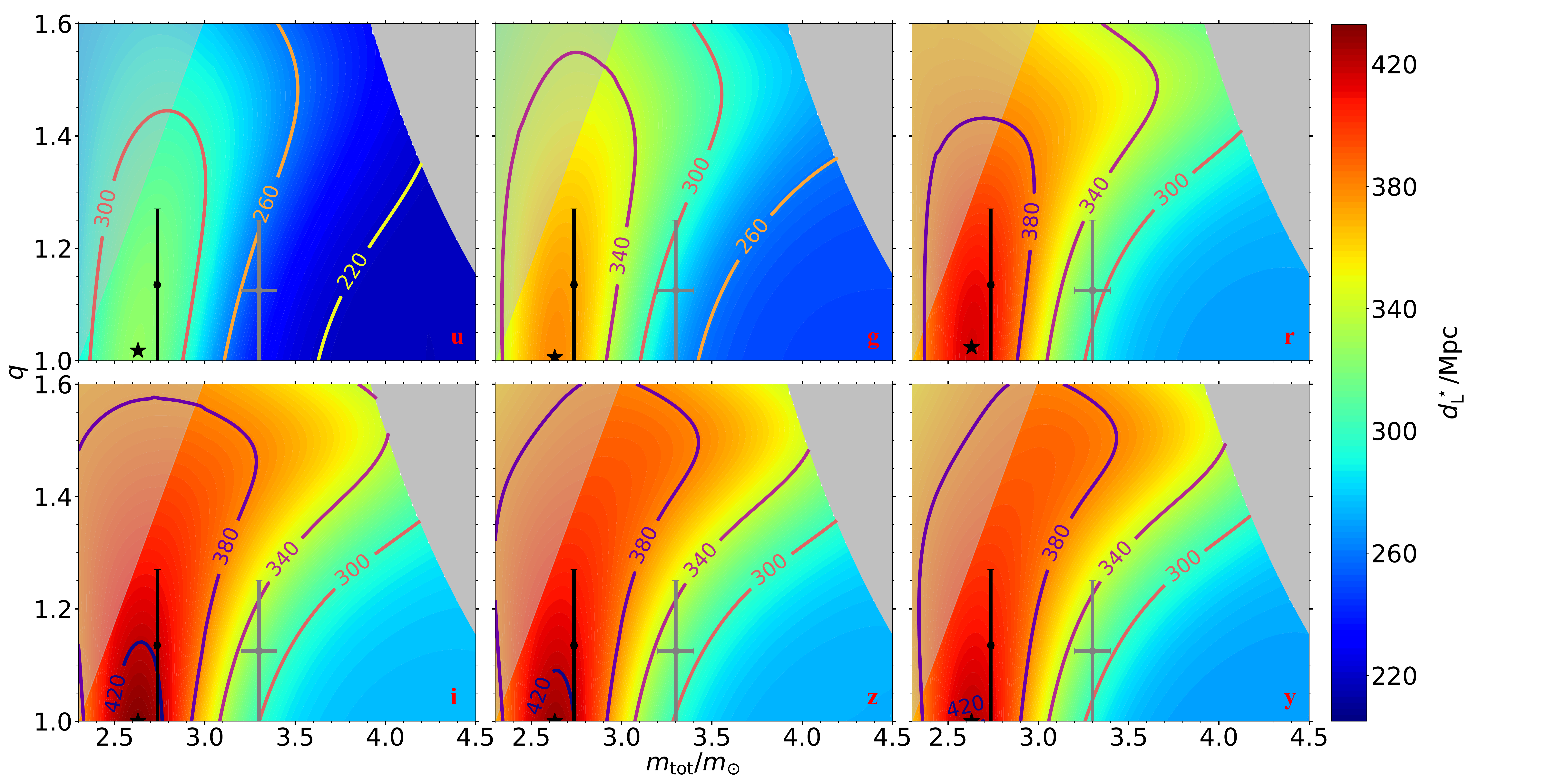}
\caption{Legends are similar to that for Fig.~\ref{fig:f5}, except that this figure shows the maximum luminosity distance ($d^{\ast}_{\rm L,max}$) for detecting those kilonovae with total mass $M_{\rm tot}$ and mass ratio $q$ at their peak luminosities, given the limiting apparent magnitude of the searching observations $m_{\rm app} = 22$. For a survey with the limiting apparent magnitude of $m_{\rm app}$, the maximum distance it can reach for a kilonova is $d_{\rm L,max}=10^{0.2(m_{\rm app}-22)}d^{\ast}_{\rm L,max}$.
}
\label{fig:f7}
\end{figure*}

\begin{figure*}
\includegraphics[width=6.5in]{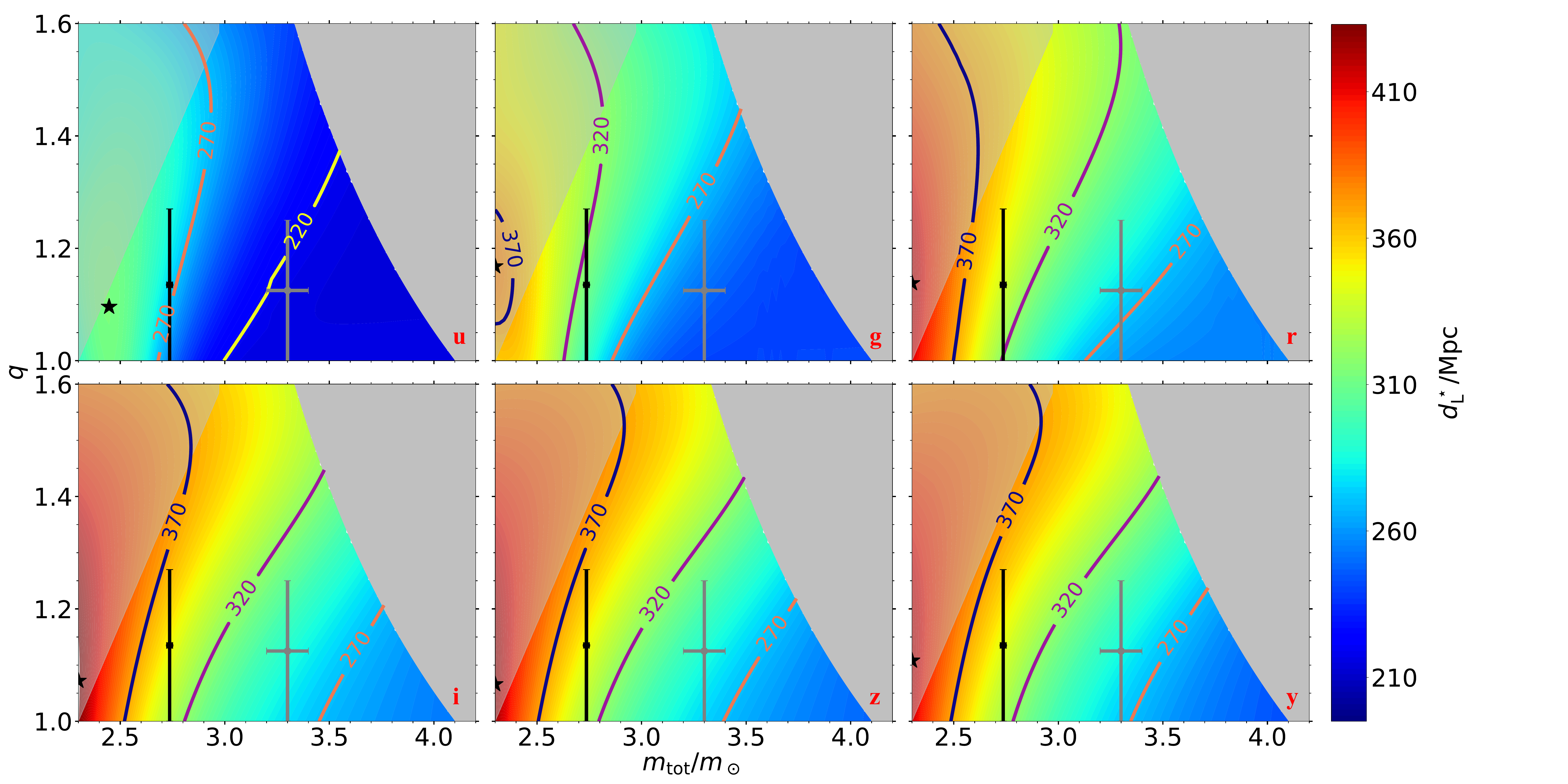}
\caption{
Legends are the same as that for  Fig.~\ref{fig:f7}, except that the EOS for neutron stars is assumed to be the SLy EOS.  }
\label{fig:f8}
\end{figure*}

\begin{figure*}
\includegraphics[scale=0.40]{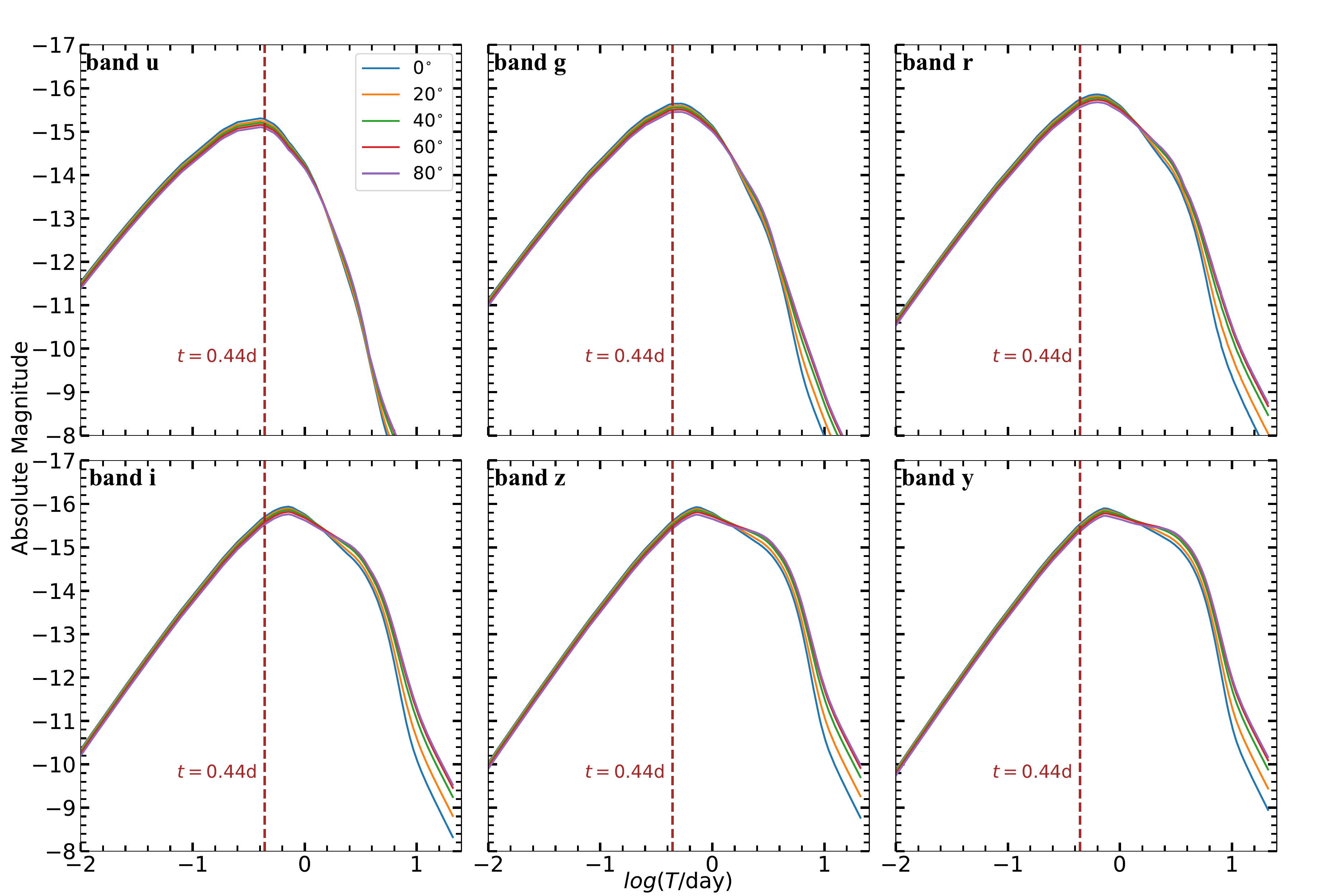}
\caption{
Kilonova LCs from GW170817-like objects viewing at $\theta_{\rm v}=0^\circ$ (blue line), $20^\circ$ (yellow line), $40^\circ$ (green line), $60^\circ$ (red line), and $80^\circ$ (purple line), respectively. Panels from left to right, from top to bottoms show the LCs for the u, g, r, i, z, and y bands, respectively. Note that only the kilonova emission is considered here though the afterglow signal could be significant when $\theta_{\rm v}$ is small (e.g., a few degree or so).
}
\label{fig:f9}
\end{figure*}

\subsubsection{Dependence of kilonova peak luminosity on BNS Parameters}
\label{subsec:Mpeak_Mtot_q}

In this subsection, we mainly focus on the kilonova peak luminosity, rather than the detailed shapes of their LCs, and its dependence on the BNS parameters and the EOS of NSs. We generate the LCs for a larger sample of BNS mergers, with different total mass $m_{\rm tot}$ and mass ratio $q$ according to our model and consequently calculate the peak luminosity for each LC. We also assume that the NS EOS is described by either the DD2 EOS or the SLy EOS as the typical representatives of the stiff or soft EOSs. For the SLy EOS or the DD2 EOS, $m_{\rm tot}$ is in the range from $2.3M_{\rm \odot}$ to $4.2M_{\rm \odot}$ or from $2.3M_{\rm \odot}$ to $4.8M_{\rm \odot}$. The minimum mass for NSs is set to be $1.15M_{\rm \odot}$ though the theoretical NS minimum mass could be much small \citep{Lattimer2012485515}. The maximum mass of non-rotating NSs can be $2.42M_\odot$ or $2.06M_\odot$ assuming the DD2 EOS or the SLy EOS. Note that the minimum and maximum masses among NSs with currently available measurements are $1.17M_{\rm \odot}$ and $2.1M_\odot$, respectively \citep[e.g.,][]{Suwa201833053312, Tauris2017170170, Cromartie20207276}.

Figures~\ref{fig:f5} and \ref{fig:f6} show the dependence of the kilonova peak luminosities at the u, g, r, i, y, and z bands on $m_{\rm tot}$ and $q$ assuming the DD2 EOS and the SLy EOS, respectively. As seen from both figures, the peak luminosities of kilonovae with fixed $q$ tend to increase with increasing $m_{\rm tot}$ when $m_{\rm tot}\lesssim 2.8M_\odot$ (or $\lesssim 2.6M_\odot$), and then decrease with increasing $m_{\rm tot}$ when $m_{\rm tot}\gtrsim 2.8M_{\rm \odot}$ (or $\gtrsim 2.6M_{\rm \odot}$) assuming the DD2 EOS (or the SLy EOS). The main reasons are as follows. In the small $m_{\rm tot}$ region, the remnant disc mass of the BNS merger is large ($\ge10^{-2}M_{\rm \odot}$) under both EOSs and saturate around $0.2M_\odot$ at the low $m_{\rm tot}$ end (large $\tilde{\Lambda}_{\rm s}$). But the dynamical ejecta mass increases with $m_{\rm tot}$ in the range from $\sim2.3M_\odot$ to $\sim2.8M_\odot$ (or from $\sim 2.3M_\odot$ to $\sim2.6M_\odot$) under the DD2 (SLy) EOS. Despite this, in the low $m_{\rm tot}$ region, the kilonova peak luminosity is dominated by the contribution from the disc wind (proportional to the remnant disc mass) and the contribution from dynamical ejecta is less significant (see Fig.~\ref{fig:f2}). Therefore, the predicted kilonova peak luminosities do not vary significantly when $m_{\rm tot}$ is small. In the high $m_{\rm tot}$ region, the merger remnant of the BNS merger with $m_{\rm tot} \gtrsim 3.2M_\odot$ (or $\gtrsim 2.9M_\odot$) under the DD2 EOS (or the SLy EOS) would promptly collapse into a BH rather than a supramassive NS (SMNS) or a hypermassive NS (HMNS) and the remnant disc mass would decrease drastically to relatively much smaller values, thus the contribution from the disc wind to the peak luminosities would be small. The emission from the dynamical ejecta may dominate the peak luminosity for a kilonova with large $m_{\rm tot}$ (see Fig.~\ref{fig:f3}), but the ejecta mass is not much larger than those with small $m_{\rm tot}$ (see Table~\ref{tab:t3}). These lead to much fainter peak luminosities for kilonovae with large $m_{\rm tot}$ compared with those with small $m_{\rm tot}$ (see also discussions in Section~\ref{subsec:PhyTempLCs}). Thus most of the predicted kilonovae with $m_{\rm tot} \gtrsim 3.5M_\odot$ have the peak luminosities fainter than $-15.5$ to $-14.7$\,mag, about $0.7-1.0$\,mag fainter than those with $m_{\rm tot} \sim 2.3-2.9M_\odot$ (see Figs.~\ref{fig:f5} and \ref{fig:f6}).

The kilonova peak luminosities may also depend on the mass ratio $q$. When $m_{\rm tot}$ is small ($\lesssim 2.9M_\odot$ for DD2 EOS or $\lesssim2.6M_{\rm \odot}$ for SLy EOS), the peak luminosities do not vary significantly with different $q$. The reason is that the remnant disc mass approaches asymptotically to $\sim 0.2M_{\rm \odot}$ and is insensitive to $q$ when $\widetilde{\Lambda}\ge750$ (i.e., small $m_{\rm tot}$) according to our settings (see Eqs.~\ref{eq:disc_rea}, and \ref{eq:disc_cf}, \ref{eq:disc}), and thus the peak luminosity is also insensitive to $q$.\footnote{However, we note that some recent NR simulations predicted that the remnant disc mass increases with increasing $q$ even when $m_{\rm tot}$ is small (e.g., $<2.9M_{\rm \odot}$) \citep[e.g.,][]{Nedora20219898}, which may result in brighter peak luminosities at large $q$ in the small $m_{\rm tot}$ region. We defer the study of the effect of this dependence to a future work by combining the NR simulation results with the semi-analytical work presented in this paper. } 
When $m_{\rm tot}$ is large, the remnant disc mass is controlled by Equation~\eqref{eq:disc_cf} and it is sensitive to $q$. As a result, the peak luminosities increase with $q$ at $m_{\rm tot} \gtrsim 2.9M_{\rm \odot}$ (or $\gtrsim 2.6M_{\rm \odot}$) for the DD2 (SLy) EOS. For example, the differences between the peak luminosities for kilonovae with $q=1$ and those with $q=1.6$ (with the same $m_{\rm tot}\gtrsim 3.5M_\odot$) can be as large as $0.6$\,mag or more (see Figs.~\ref{fig:f5} and \ref{fig:f6}).

In general, the distributions of the peak luminosities on the $m_{\rm tot}-q$ plane resulting from different choices of the EOS are similar, but there are still somewhat differences. For example, the brightest kilonova can be produced by BNS mergers with $m_{\rm tot} \sim 2.6-2.75M_\odot$ and $q\sim 1.0-1.2$ if adopting the DD2 EOS, while it is by BNS mergers with $m_{\rm tot} \sim 2.3-2.5M_\odot$ and $q\sim 1.0-1.25$ (at u, g, r, i, z, and y bands) if adopting the SLy EOS. The kilonovae generated from BNS mergers by adopting the DD2 EOS are systematically brighter than those from adopting the SLy EOS by $\sim0.2-0.5$\,mag. As seen from Figures~\ref{fig:f5} and \ref{fig:f6}, the mean peak magnitudes at different bands estimated for BNS mergers with parameters similar to GW170817 are fainter than those of GW170817 if adopting the SLy EOS, while they are more or less the same as GW170817 observations if adopting the DD2 EOS. This may suggest that GW170817 may be a kilonova event significantly brighter than its parent population with the same $m_{\rm tot}$ and $q$ under the SLy EOS while it is the typical case  under the DD2 EOS.  

Note here that kilonovae resulting from BNS mergers with the same $m_{\rm tot}$ and $q$ may not have the same peak luminosities as the masses of the dynamical ejecta and remnant disc given by different NR simulations have some scatters. Figures~\ref{fig:f5} and \ref{fig:f6} shown here are obtained by smoothing the corresponding scatters in the resulting kilonova peak luminosities and only represent the average ones (see also Section~\ref{subsec:parasettings}). 

Figures~\ref{fig:f7} and \ref{fig:f8} show the maximum distances that the kilonovae from BNS mergers with $m_{\rm tot}$ and $q$ may be detectable at their peak luminosity by a telescope with the limiting magnitude of $m_{\rm app}=22$ in the u, g, r, i, z, or y band. These two figures are similar to Figures~\ref{fig:f5} and \ref{fig:f6} but with a conversion from the peak magnitudes to the maximum detectable distances for any telescope with a limiting magnitude of $m_{\rm app}$ at different bands. As clearly seen from these figures, with the same limiting magnitude $m_{\rm app}$, one can detect kilonovae at larger distances by adopting the red filters (e.g., the H-band shown by the two bottom right panels) than that by adopting the blue filters (e.g., the u band shown by the two top left panels). Kilonovae from BNS mergers with $m_{\rm tot}$ around $2.7-2.9M_\odot$ and small $q$ (close to 1) may be detectable by telescopes within a distance of $(260-400)\times 10^{0.2(22-m_{\rm app})}$\,Mpc (from u  to H bands), while those with $m_{\rm tot} \gtrsim 3.5M_\odot$ and small $q$ (close to $1$) may be only detectable within a distance of $(210-300)\times 10^{0.2(22-m_{\rm app})}$\,Mpc.

The kilonova peak luminosities and LCs also depend on the viewing angle ($\theta_{\rm v}$). Figure~\ref{fig:f9} shows the LCs for kilonovae with intrinsic parameters the same as GW170817 but different $\theta_{\rm v}$ at the u, g, r, i, z, and y bands. Apparently, the peak luminosities decrease with increasing $\theta_{\rm v}$ and the differences between the peak magnitudes obtained by assuming $\theta_{\rm v}\sim 0^\circ-20^\circ$ and $80^\circ$ are $\sim 0.4$\,mag in different bands. The slightly high difference in the u band is due to that the u band emission is mainly contributed by the blue component, more anisotropic comparing with the red component, which may dominate the emission in the y-band. Furthermore, the shapes of the LCs also depend on $\theta_{\rm v}$ and they are more extended for those with large $\theta_{\rm v}$. As seen from Fig.~\ref{fig:f9}, the red ``bump''s of the kilonova LCs in the red bands (e.g., z- and y-band) at $T=2$\,day become more evident when $\theta_{\rm v}$ is large (e.g., $80^\circ$) as the red component is distributed more closer to the equatorial region.

LIGO/Virgo/KAGRA (LVK) will have O4 observation in 2023 and the median detectable BNS distance can be up to $352.8^{+10.3}_{-9.8}$\,Mpc \citep{Petrov20225454}. According to Figures~\ref{fig:f7} and \ref{fig:f8}, it is highly like to detect the optical EM counterparts for almost all GW170817-like BNS mergers observed by O4 if the limiting magnitude of the survey searching for the associated kilonovae can be $\sim22$\,mag, but the kilonovae of which the BNS properties similar to GW190425 need to be searched with a limiting magnitude fainter than $\sim23-24$\,mag. O5 observation may start later and end in 2028. At that time, it can reach a median distance of $620^{+16}_{-17}$\,Mpc for BNS mergers \citep{Petrov20225454}, therefore, probably almost all GW170817-like BNS mergers must be detected by surveys with limiting magnitude fainter than $\sim23$\,mag. However, those with properties like GW190425 or with $M_{\rm tot} \gtrsim 3.5M_\odot$ may be hard to be detected if searching magnitude is not fainter than $24-25$\,mag.

\section{Kilonova Luminosity Functions}
\label{sec:LF}

One may obtain the KLFs after detecting a large number of kilonovae. We are going to model the luminosity functions theoretically using our semi-analytical model and the population synthesis model for comparison with the future observations. The KLFs reflect both the cosmic generation rate of kilonovae as a population and the underlying physical processes that govern the kilonova phenomena. In principle, the KLFs per unit time\footnote{Here the luminosity function per unit time is defined because the kilonovae are transient events and they disappear in a relatively short time after their emergent.} can be estimated as
\begin{eqnarray}
\frac{d\Phi (M_{\rm ab},z)}{dt_{\rm obs}} & = & \frac{1}{1+z} 
\iiint dm_{\rm tot} dq d\theta_{\rm v} P(m_{\rm tot}|z)  P(q|m_{\rm tot},z)  \nonumber \\
& & \times P(\theta_{\rm v}) P(M_{\rm ab}|m_{\rm tot},q,\theta_{\rm v})  \frac{d^2N(z)}{dtdV_{\rm c}},  
\label{eq:lum_func}
\end{eqnarray}
at any specific band with the peak magnitude of $M_{\rm ab}$. Here $d\Phi(M_{\rm ab},z)/dt_{\rm obs}$ is defined as the comoving number density of kilonovae with absolute peak magnitude at a given band in the range from $M_{\rm ab} + dM_{\rm ab}$ in a unit time period from $t_{\rm obs}$ to $t_{\rm obs} + dt_{\rm obs}$, and it has the unit of ${\rm Gpc}^{-3}{\rm mag}^{-1}{\rm yr}^{-1}$. One may also derive the KLFs in the rest frame of kilonovae rather than the observers' rest frame, i.e., $d\Phi(M_{\rm ab},z)/dt = (1+z) d\Phi(M_{\rm ab},z)/dt_{\rm obs}$. In the above equation, $P(m_{\rm tot}|z)$, $P(q|m_{\rm tot},z)$, $P(\theta_{\rm v})$, and $P(M_{\rm ab}|m_{\rm tot},q,\theta_{\rm v})$ are the probability distributions of the total mass, mass ratio, and viewing angle of BNS mergers at redshift $z$, and the probability distribution of the peak magnitude at a given band for BNS mergers with given $m_{\rm tot}$, $q$, and $\theta_{\rm v}$\footnote{Note that the peak magnitude here is a distribution for a given set of ($m_{\rm tot},q,\theta_{\rm v}$), which is different from the mean one shown in Figs.~\ref{fig:f5} and \ref{fig:f6}.}. In our model, the minimum mass of the NSs is assumed to be $1.15M_{\rm \odot}$ as discussed in Section~\ref{subsec:Mpeak_Mtot_q}. The quantity $d^2N(z)/dtdV_{\rm c}$ denotes the comoving number density of BNS mergers occurred at redshift $z$ in the cosmic time range from $t$ to $t+dt$ and $dt = dt_{\rm obs}/(1+z)$.

To estimate the KLFs, one needs to first obtain all the three probability distribution functions $P(M_{\rm ab}|m_{\rm tot},q,\theta_{\rm v})$, $P(m_{\rm tot}|z)$, $P(q|m_{\rm tot},z)$, $P(\theta_{\rm v})$, and also $d^2 N(z)/dtdV_{\rm c}$ in the above equation. We obtain $P\left(M_{\rm ab}|m_{\rm tot},q,\theta_{\rm v}\right)$ according to our semi-analytical kilonova model introduced in Sections~\ref{sec:model} and \ref{sec:gkilonova}. $d^2N(z)/dtdV_{\rm c}$, $P(m_{\rm tot}|z)$, and  $P(q|m_{\rm tot},z)$ can be obtained by combining the population synthesis model for binary stars with the cosmic star formation history and metallicity evolution. In this paper, we adopt the binary star evolution (BSE) model `$\alpha$10bk$\beta$0.9' given in \citet{Chu202215571586}, which can match both the Galactic BNS observations and the BNS merger rate constrained by the GW observations, and we adopt the cosmic star formation rate given in \citet{Madau2014415486} and the metallicity evolution in \citet{Belczynski2016512515}. We further consider the effect of EOS on the distributions of $m_{\rm tot}$ and $q$. In our BSE model, we assume a linear relation between the NS mass $m_{\rm NS}$ and the CO-core mass at the time of supernova explosion $m_{\rm c,SN}$, for simplicity \citep[][]{Hurley2000543569}. We assume: (i) if $1.6 M_{\odot}<m_{\rm c,SN}<2.25M_{\odot}$, the star becomes an electron-degenerate oxygen-neon white dwarf, which may become an NS via the accretion-induced collapse; (ii) if $m_{\rm c,SN}\geq 2.25M_{\odot}$, the star becomes an NS through supernova explosion; (iii) if $m_{\rm c,SN}>7M_{\odot}$, the star will most likely become a black hole \citep[][]{Hurley2000543569}. In the simulations, we set a minimum NS mass of 1.15$M_{\odot}$ \citep[][]{Suwa201833053312} and the maximum NS mass of $M_{\rm TOV}$ inferred for each EOS being considered and the observations of GW170817 \citep[][]{Chabanat1998231256, Douchin2001151167, Typel20101580315803}. We ignore the NS spin in our BSE model, but note that the NS masses could extend to the mass range larger than $M_{\rm TOV}$ if considering rapid spinning NSs \citep[e.g.,][]{Douchin2001151167}.

\begin{figure*}
\includegraphics[width=6.5in]{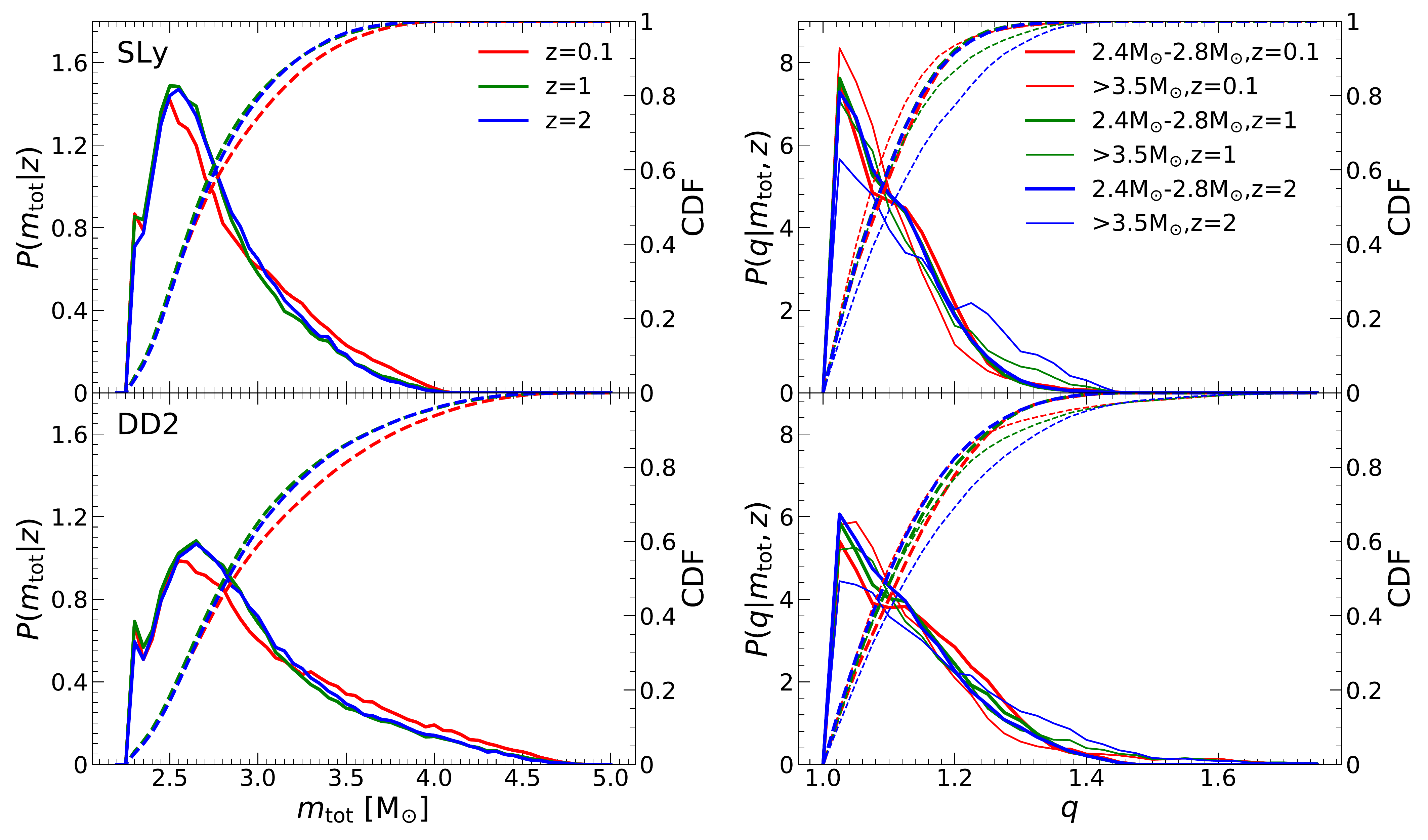}
\caption{
Distributions of the total mass ($m_{\rm tot}$; left-hand panels) and the mass ratio ($q$; right-hand panels) of merging BNSs at different redshift for the SLy EOS model (top panels) and the DD2 EOS model (bottom panels), respectively. In each panel, the red, green, and blue solid/dashed lines represent the differential/cumulative distributions of $m_{\rm tot}$ and $q$ at $z=0.1$, $1$, and $2$, respectively. In the right-hand panels, the thick and thin lines represent the cases for those with $m_{\rm tot} \in (2.4M_\odot,2.8M_\odot)$ and $m_{\rm tot}>3.5M_\odot$, respectively.
}
\label{fig:f10}
\end{figure*}

Figure~\ref{fig:f10} shows the total mass and mass ratio distributions of BNS mergers occurred at different redshift by assuming either the DD2 EOS or the SLy EOS. As seen from this figure, the predicted $P(m_{\rm tot}|z)$ peaks at $\sim 2.3-2.4M_\odot$ if adopting the SLy EOS, slightly smaller than that obtained from the case adopting the DD2 EOS (peaking at $\sim 2.3-2.6M_\odot$). As expected, adopting the DD2 EOS results in many more massive BNSs (with $m_{\rm tot} \geq 3.5M_\odot$) comparing with the case adopting the SLy EOS. Under the DD2 EOS, the fraction of BNSs with $m_{\rm tot}\geq 3.5M_\odot$ is $\sim 14-19\%$, while it is only $\lesssim 3-6\%$ under the SLy EOS. The total mass distribution is only weakly dependent on redshift whether adopting the DD2 EOS or the SLy EOS, and it is skewed towards more massive BNSs at lower redshift. The BNS mass ratio distribution is concentrated at small values, i.e., close to symmetric binaries (left-hand panels of Fig.~\ref{fig:f10}). For those BNSs around the peak of the mass distribution, with $m_{\rm tot} \in (2.3M_\odot,2.8M_\odot)$, the fraction of BNSs with $q \in (1,1.15)$ is $\sim 70\%$ (or $\sim 54\% $) at $z=0.1$ in the case adopting the SLy EOS (or the DD2 EOS). For those at the high-mass tail, e.g., with $m_{\rm tot}>3.5M_\odot$, the fraction of BNSs with $q \in (1,1.15)$ is slightly larger, i.e., $\sim 78\%$ (or $\sim 62\% $) at $z=0.1$ in the case adopting the SLy EOS (or the DD2 EOS). The mass ratio distribution is also weakly dependent on the redshift. For BNSs with high mass (e.g., $m_{\rm tot}>3.5M_\odot$), the larger the redshift, the larger the fraction of BNSs with higher $q$, while for BNSs with $m_{\rm tot}$ around the peak, the redshift dependence of the mass ratio distribution appears much weaker. 

\begin{figure*}
\includegraphics[width=6.5in]{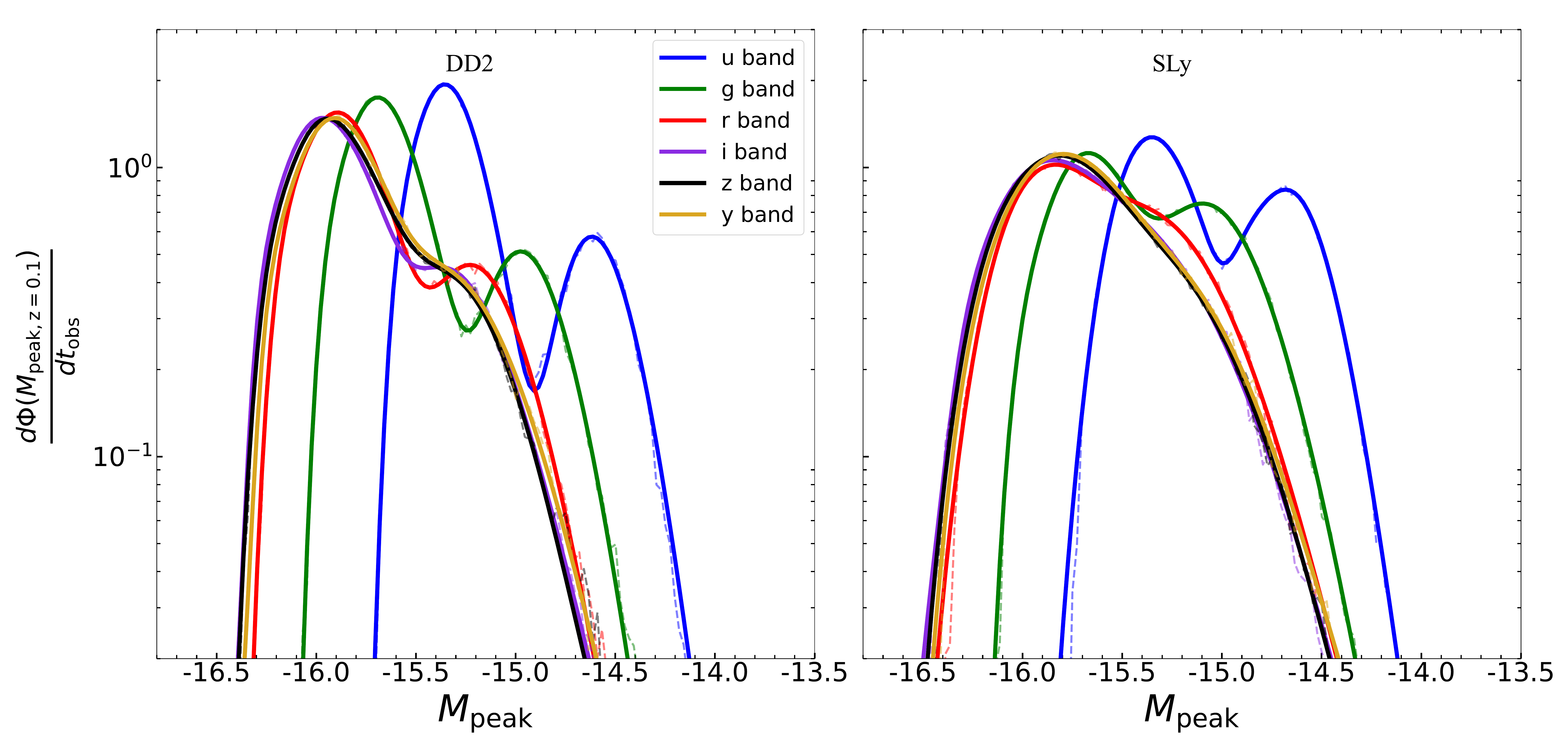}
\caption{The normalized luminosity functions estimated for the kilonovae from mergers of binary neutron stars at redshift $z\lesssim 0.1$ ``detected'' by GW detectors by assuming either the DD2 EOS (left-hand panel) or the SLy EOS (right-hand panel). The blue, green, red, purple, black, and brown dotted curves represent the normalized KLFs obtained from the Monte Carlo calculations for the u, g, r, i, z, and y bands, respectively. The blue, green, red, purple, black, and brown solid lines represent the best fits to the corresponding dotted curves, respectively.
}
\label{fig:f11}
\end{figure*}

For all BNS mergers, their viewing angle should be randomly distributed, i.e., $P(\theta_{\rm v}) \propto \sin \theta_{\rm v}$. Considering the GW signal detection, however, $P(\theta_{\rm v})$ is given by Equation~\eqref{eq:view angle}, and one can detect more BNS mergers at some preferred viewing angles. Therefore, the nKLFs obtained from observations may be modulated by this GW selection effect and different from the underlying intrinsic ones. Below, we estimate both the intrinsic nKLFs and those directly obtained from observations by the GW selection on the viewing angles. 

The KLFs evolve with redshift because not only the merger rate density but also the distributions of the properties of BNS merger systems ($m_{\rm tot}$ and $q$) may evolve with redshift. If the distributions of BNS merger properties only weakly or even do not evolve with redshift, then the KLF redshift evolution is mainly determined by the BNS merger rate density evolution, which simply induces a normalization change of the luminosity function. For simplicity, this redshift-dependent normalization can be removed by defining a normalized KLF (hereafter nKLF) as
\begin{eqnarray}
\phi(M_{\rm ab},z) & \equiv & \frac{\frac{d\Phi (M_{\rm ab},z)}{dt_{\rm obs}}}{\int d M_{\rm ab} \frac{d\Phi (M_{\rm ab},z)}{dt_{\rm obs}}} \nonumber \\
& = & \iiint dm_{\rm tot} dq d\theta_{\rm v} P(m_{\rm tot}|z)  P(q|m_{\rm tot},z)  \nonumber \\
& & \times P(\theta_{\rm v}) P(M_{\rm ab}|m_{\rm tot},q,\theta_{\rm v}).
\label{eq:normal_lf}
\end{eqnarray}
According to the above settings, we can generate a large mock sample for kilonovae and obtain the nKLFs. We also fit the obtained nKLFs by using the following form, i.e., the summation of multiple Schechter functions, 
\begin{equation}
\begin{aligned}
\phi_{\rm i} \left(M_{\rm ab}\right) = \sum^n_{i=1} \phi^\ast_{\rm i}10^{0.4b_{\rm i}(M^\ast_{\rm i} - M_{\rm ab})} e^{-10^{0.4a_{\rm i}(M^\ast_{\rm i}-M_{\rm ab})}},
\label{eq:fit_lum_func}
\end{aligned}
\end{equation}
where we adopt either $n=3$ to fit the nKLFs. The best-fitting parameters are listed in Table~\ref{tab:t4}. 

Figure~\ref{fig:f11} shows the resulting multiband nKLFs [i.e., $\phi(M_{\rm  ab},z)$] for BNS mergers detected by GW detectors at redshift $z\lesssim0.1$ under the assumption of two different EOSs (i.e., DD2 and SLy) and it also shows the best fits to the resulting nKLFs. As seen from this figure, the nKLFs in the u, g, r, i, z, and y bands appears to be bimodal, with the bright peaks at $\sim -15.36$, $-15.68$, $-15.89$, $-15.95$, $-15.95$ and $-15.91$\,mag (or $\sim -15.34$, $-15.66$, $-15.84$, $-15.86$, $-15.82$, and $-15.80$\,mag), respectively, and the faint peaks at $\sim -14.61$, $-14.98$, $-15.20$, $-15.29$, $-15.25$ and $-15.32$\,mag (or $\sim -14.61$, $-15.04$, $-15.25$, $-15.32$, $-15.14$ and $-15.11$\,mag), respectively, assuming the DD2 EOS (or the SLy EOS). The bright peaks of nKLFs in the redder bands usually have relatively smaller magnitudes (thus brighter) than those in the bluer bands, while the faint peaks of nKLFs in the redder bands usually have relatively larger magnitudes thus fainter than those in the bluer bands. The nKLFs in the redder bands appear to be wider than those in the bluer bands. 

The nKLF at each band has a clear cut-off at the bright end as shown in Figure~\ref{fig:f11}. The nKLFs for the r, i, z, and y bands have the bright end cut-off at $\left[-16.4,-16.3\right]$/$\left[-16.4,-16.3\right]$\,mag, while those for the g  and u bands at $\sim-16.1$/$-16.1$\,mag  and  $\sim-15.7$/$-15.7$\,mag, respectively, if adopting the DD2/SLy EOS. The bright end cut-off is due to that the remnant disc mass asymptotically converges to $m_{\rm disc}\sim 0.2M_{\rm \odot}$ at the high $m_{\rm disc}$ end (see Section~\ref{subsec:Mpeak_Mtot_q}), usually resulting from BNS mergers with small $m_{\rm tot}$, for which the wind ejectas from the discs dominate the kilonova luminosities.

\begin{figure*}
\includegraphics[width=6.5in]{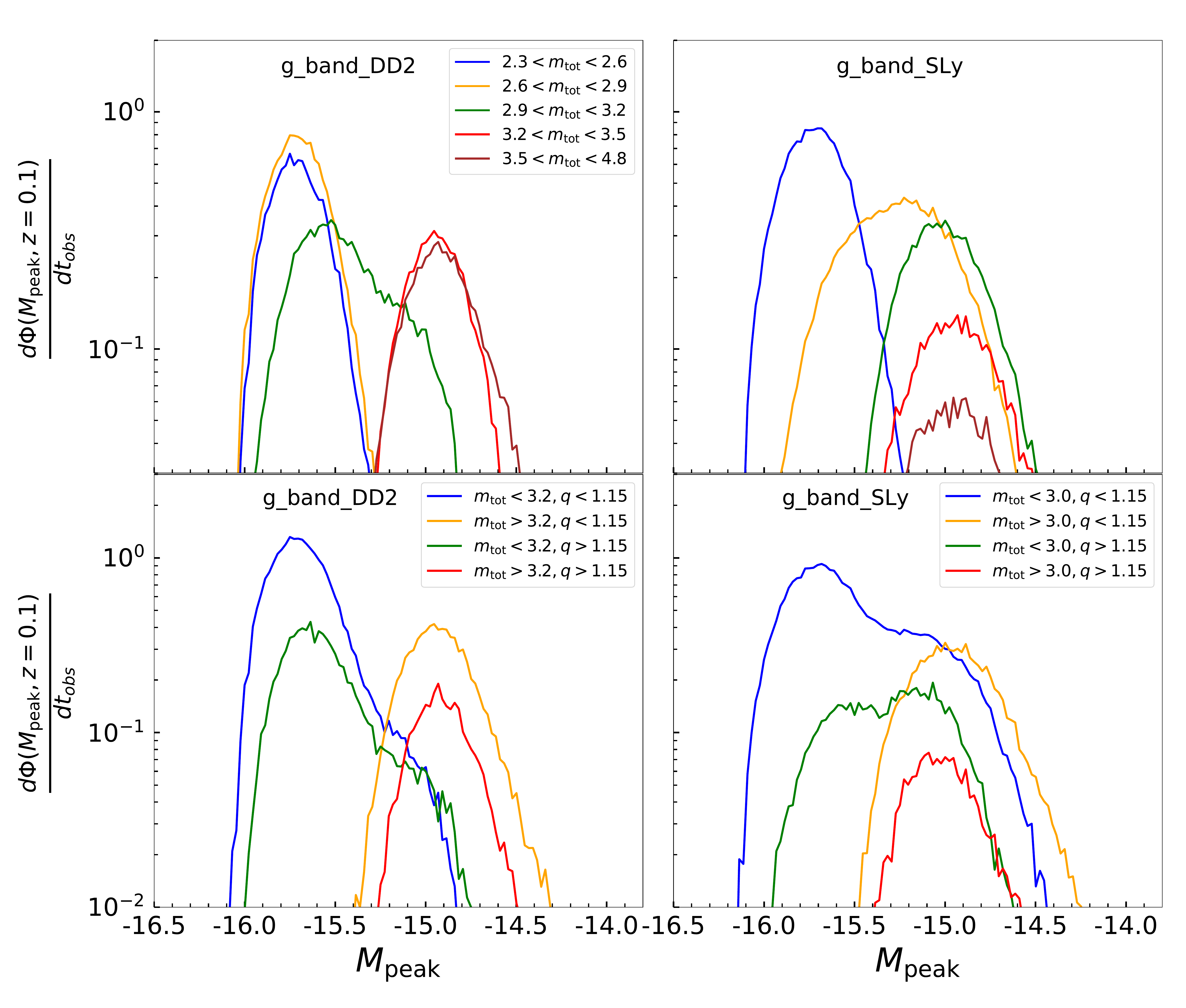}

\caption{
Contributions from BNSs with different total mass $m_{\rm tot}$ (top panels) and mass ratio $q$ (bottom panels) to the total kilonova g-band luminosity function (see Fig.~\ref{fig:f11}) under the assumption of either the DD2 EOS (left-hand panels) or the SLy EOS (right-hand panels). In the top two panels, blue, yellow, green, red, and dark brown dashed curves show the contributions from those BNS mergers with $m_{\rm tot} \in [(2.4M_\odot,2.7M_\odot)$, $[(2.7M_\odot,3.0M_\odot)$, $[(3.0M_\odot,3.3M_\odot)$, $[(3.3M_\odot,3.6M_\odot)$, and $[(3.6M_\odot,4.4M_\odot)$, respectively. In the bottom two panels, blue, yellow, green, and red dashed curves show the contributions from those BNS mergers with ($m_{\rm tot} <3.2M_\odot$, $q<1.15$),  ($m_{\rm tot} \ge 3.2M_\odot$, $q<1.15$), ($m_{\rm tot} <3.2M_\odot$, $q\ge1.15$), and ($m_{\rm tot} \ge 3.2M_\odot$, $q \ge 1.15$), respectively. 
}
\label{fig:f12}
\end{figure*}

The overall shapes of the nKLFs are controlled by both the distribution of BNS properties and the distribution of the absolute peak magnitude of kilonovae resulting from the BNS mergers with given properties (see equation~\ref{eq:normal_lf}), therefore it may be understood by isolating the contributions from BNS mergers with different $m_{\rm tot}$ and $q$ to the nKLFs, for example, as shown in Figure~\ref{fig:f12} for the total g-band nKLF (see the green curves in Fig.~\ref{fig:f11}). As seen from this figure, the faint and bright components of the nKLF obtained under the DD2 EOS are mainly contributed by those BNS mergers with $m_{\rm tot} >3.2M_\odot$ and $m_{\rm tot}\lesssim 3.2M_\odot$, respectively, and neither those BNS mergers with small $m_{\rm tot}$ nor those with large $m_{\rm tot}$ contribute much to the nKLF at $M_{\rm ab} \sim -15.2$. First, for the BNS mergers with $m_{\rm tot} <3.1M_\odot$, as seen from Figure~\ref{fig:f5}, the peak luminosities that result from different BNS mergers are bright ($\lesssim -15.5$) and the differences in the peak luminosities from different mergers are small. Among them, those with $m_{\rm tot}\lesssim 2.9M_{\rm \odot}$ result in kilonovae with the peak magnitudes brighter than $\sim-15.7$\,mag because large remnant discs are formed ($m_{\rm disc}>0.1M_\odot$) during the merging processes and the kilonova luminosities are mainly contributed by the disc winds. 
Second, for the BNS mergers with $m_{\rm tot} >3.4M_\odot$, the peak luminosities result from different BNS mergers are faint ($\lesssim -15.1$) and the differences in the peak luminosities from different mergers are also small. For these mergers, the remnant discs have negligible mass and thus contribute little to the kilonova luminosities, while the dynamical ejecta have the mass in a narrow range of $\sim10^{-3}-10^{-2}M_{\rm \odot}$ and it dominates the kilonova peak luminosities. Last, for those BNS mergers with $m_{\rm tot} \sim 3.1-3.4M_\odot$, the peak luminosities decrease quite sharply with increasing $m_{\rm tot}$. These are the primary reasons for the bimodal distribution of the g-band nKLF. For the nKLFs in the other bands, the reason for the bimodal distribution is more or less the same. We note here that \citet{Setzer2022220512286} recently estimated the KLFs and found the bimodal structure, similar to our results. The peak magnitudes of the nKLF faint components we obtained are relatively brighter than those obtained by \citet{Setzer2022220512286} because the heating from the fallback accretion is considered in our model but not in \citet{Setzer2022220512286}.

For the nKLFs obtained for the u, g, r, i, z, or y band, the bright component contributes the majority of kilonovae, i.e., $77.1\%$, $78.1\%$, $78.0\%$, $80.8\%$, $89.2\%$, or $77.2\%$ for the DD2 EOS ($63.0\%$, $62.6\%$, $75.3\%$, $77.7\%$, $91.9\%$, $92.0\%$ for the SLy EOS), to the corresponding nKLFs, while the faint components contribute a fraction of $22.9\%$, $21.9\%$, $22.0\%$, $19.2\%$, $10.8\%$, or $22.8\%$ for the DD2 EOS ($37.0\%$, $37.4\%$, $24.7\%$, $22.3\%$, $8.1\%$, $8.0\%$ for the SLy EOS) to the corresponding nKLFs. Obviously, the fraction of kilonova in the nKLF faint component is significantly smaller than that in the bright component. The primary reason is that the low-mass ($m_{\rm tot}<2.9M_{\rm \odot}$/$2.8M_{\rm \odot}$ for the DD2/SLy EOS) and the more symmetric ($q<1.18$/$1.12$ for the DD2/SLy EOS) mergers dominate the total distribution $P(m_{\rm tot}|z)$ and $P(q|m_{\rm tot},z)$ (see Fig.~\ref{fig:f10}), which contribute mostly to the bright component of the nKLFs as discussed above. In addition, the heights of the nKLF bright peaks are $\sim 4$ times higher than those of the faint peaks under the DD2 EOS (see left-hand panel in Fig.~\ref{fig:f11}), while the heights of the nKLF bright peaks are $\lesssim 2$ times higher than those of the faint peaks under the SLy EOS.

Despite that the general shapes of the nKLFs resulting by assuming the DD2 EOS are indeed different from those by assuming the SLy EOS, the bright-end cut-offs, the peaks of both the bright and faint components for the nKLF at each given band obtained by assuming the DD2 EOS are more or less the same as those obtained by assuming the SLy EOS, respectively, and the differences are $\lesssim 0.1-0.2$\,mag (see Fig.~\ref{fig:f11}). The reason is that  the nKLFs are controlled by both the distributions of the BNS properties ($m_{\rm tot}$ and $q$) and the kilonova peak magnitude distribution (at any given band) for BNS mergers with given properties. Although the kilonova peak magnitude distributions for a given set of BNS properties obtained under the DD2 EOS are quite different from those obtained under the SLy EOS as seen from Figures~\ref{fig:f5} and \ref{fig:f6}. For example,  for a BNS merger with $m_{\rm tot}\sim 3.0M_\odot$ and $q=1$, the kilonova magnitude at the g band is $\sim -15.7$ or $\sim -15.4$ by assuming the DD2 EOS or the SLy EOS. However, the total mass distribution $P(m_{\rm tot}|z)$ obtained under the DD2 EOS peaks at higher $m_{\rm tot}$ ($\sim 2.65M_\odot$) than that obtained under the SLy EOS ($\sim 2.55$), thus there are relatively more lower mass BNS mergers under the SLy EOS than that under the DD2 EOS (see Fig.~\ref{fig:f10}). As the kilonova with lower mass is brighter than that with higher mass, therefore, the effects of $P(m_{\rm tot}|z)$ and $P(q|m_{\rm tot},z)$ partly cancel the effects of $P(M_{\rm ab}|m_{\rm tot},q,\theta_{\rm v})$ on the nKLFs, and thus lead to smaller differences between the bright-end cut-offs, peaks positions of both the bright and faint components of the nKLFs obtained from different EOSs.  

\begin{figure*}
\includegraphics[width=6.5in]{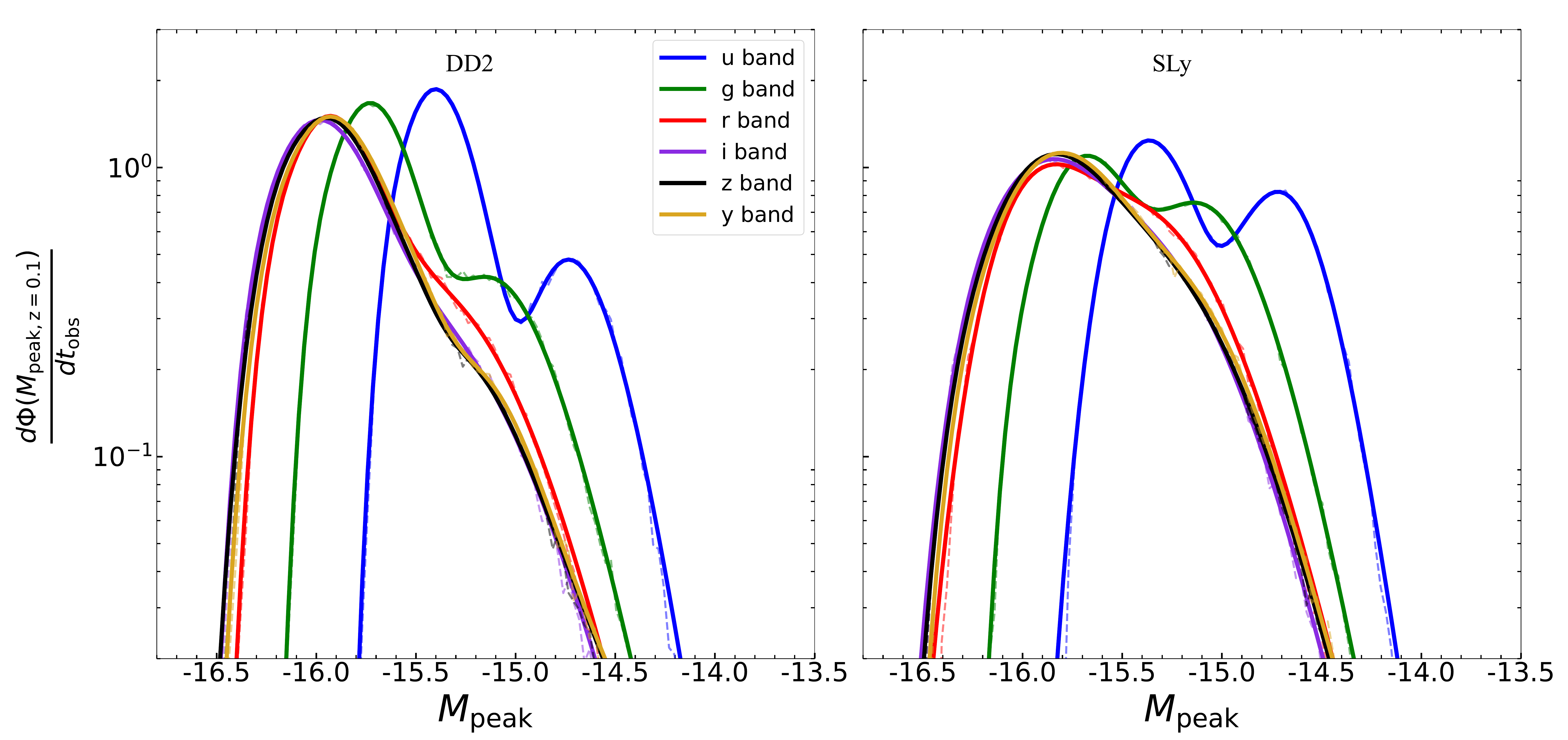}
\caption{
Legend similar to that for Fig.~\ref{fig:f11}, except that the selection effect by the GW detection on the viewing angle is ignored. 
}
\label{fig:f13}
\end{figure*}

Figure~\ref{fig:f13} shows the obtained multiband nKLFs ($\phi(M_{\rm  ab},z)$) at redshift $z=0.1$ with viewing angles of BNSs randomly distributed, (i.e., not affected by the selection effects for kilonovae associated with the BNS mergers detected by GW detectors). By comparing this figure with Figure~\ref{fig:f11}, it can be seen that the double-peak structures are less sharp in the case with randomly distributed viewing angles. The main reason is that there are many more BNS mergers with large $\theta_{\rm v}$ in this case, which produce relatively fainter kilonovae (see Fig.~\ref{fig:f9}) and thus lead to the broadening of both the bright and faint components of the nKLFs, while there are little BNS mergers with $\theta_{\rm v}$ close to $\pi/2$ in the case with GW selected viewing angle distribution. The best-fitting parameters of these nKLFs are listed in Table~\ref{tab:t5}.

\begin{table*}
\centering
\caption{The fitting coefficients of the anisotropic luminosity functions (expressed as the equation \ref{eq:fit_lum_func}) of different kilonovae samples grouped according to the total mass, mass ratio and EOS.}
\begin{tabular}{|c|c|c|c|c|c|c|c|c|c|c|c|c|c|} \hline \hline
EOS & Band & $\phi_1$ & $\phi_2$ & $\phi_3$ & $a_1$ & $a_2$ & $a_3$ & $b_1$ & $b_2$ & $b_3$ & $M_1$ & $M_2$ & $M_3$ \\ \hline
\multirow{6}{*}{DD2} & u & 1.28 & $1.03$ & $1.04$$\times$$10^{-3}$ & 9.49 & 1.41 & 4.31 & 13.88 & 2.63 & 17.99 & -15.48 & -15.04 & -13.82 \\ \cline{2-14}
 & g & $1.76$ & $3.12$ & $1.65$$\times$$10^{-2}$ & 10.28 & 9.89 & 5.35 & 9.70 & 2.77 & 14.44 & -15.87 & -15.54 & -14.32 \\ \cline{2-14}
 & r & $1.56$ & $1.39$ & $3.93$$\times$$10^{-2}$ & 10.93 & 3.99 & 4.31 & 10.32 & 2.73 & 12.34 & -16.15 & -15.63 & -14.59 \\ \cline{2-14}
 & i & 1.27 & $2.60$ & $3.95$$\times$$10^{-1}$ & 12.43 & 3.24 & 5.42 & 7.03 & 3.73 & 9.13 & -16.25 & -15.88 & -15.02 \\ \cline{2-14}
 & z & 3.24 & $5.02$$\times$$10^{-1}$ & $2.85$$\times$$10^{-4}$ & 7.95 & 4.46 & 5.72 & 17.04 & 2.19 & 15.38 & -15.98 & -15.95 & -14.28 \\ \cline{2-14}
 & y & $7.22\times10^{-1}$ & $2.26\times10^{-1}$ & $1.81$$\times$$10^{-3}$ & 11.43 & 11.18 & 2.74 & 11.57 & 1.79 & 11.27 & -16.17 & -15.33 & -14.18 \\  \hline
\multirow{6}{*}{SLy} & u & $5.5$$\times$$10^{-2}$ & $4.20$$\times$$10^{-1}$ & $1.68$$\times$$10^{-1}$ & 2.49 & 12.61 & 4.75 & 13.77 & 3.93 & 14.61 & -14.64 & -14.56 & -14.23 \\ \cline{2-14}
 & g & $5.62$$\times$$10^{-1}$ & $1.53$ & $1.96$$\times$$10^{-3}$ & 8.00 & 11.13 & 3.55 & 8.29 & 1.90 & 12.49 & -15.89 & -15.40 & -13.97 \\ \cline{2-14}
 & r & $1.94$ & $2.81\times10^{-3}$ & $2.12$$\times$$10^{-5}$ & 2.82 & 5.48 & 1.67 & 10.21 & 3.22 & 10.66 & -15.59 & -14.06 & -12.25 \\ \cline{2-14}
 & i & $1.34$ & $9.71\times10^{-2}$ & $4.96$$\times$$10^{-2}$ & 4.02 & 5.17 & 1.70 & 7.34 & 2.25 & 8.56 & -15.96 & -14.78 & -14.53 \\ \cline{2-14}
 & z & $1.25$ & $2.84\times10^{-1}$ & $2.12$$\times$$10^{-2}$ & 3.92 & 5.16 & 1.96 & 7.07 & 2.12 & 9.30 & -15.94 & -15.01 & -14.37 \\ \cline{2-14}
 & y & $9.45\times10^{-1}$ & $7.23\times10^{-1}$ & $1.55$$\times$$10^{-2}$ & 4.58 & 4.29 & 2.14 & 6.22 & 2.08 & 9.40 & -16.03 & -15.19 & -14.32 \\ \hline

\end{tabular}
\label{tab:t4}
\end{table*}

\begin{table*}
\centering
\caption{The fitting coefficients of the isotropic luminosity functions (expressed as the equation \ref{eq:fit_lum_func}) of different kilonovae samples grouped according to the total mass, mass ratio and EOS.
}
\begin{tabular}{|c|c|c|c|c|c|c|c|c|c|c|c|c|c|} \hline \hline
EOS & Band & $\phi_1$ & $\phi_2$ & $\phi_3$ & $a_1$ & $a_2$ & $a_3$ & $b_1$ & $b_2$ & $b_3$ & $M_1$ & $M_2$ & $M_3$ \\ \hline
\multirow{6}{*}{DD2} & u & $1.47$ & $3.95$$\times$$10^{-1}$ & $3.05$$\times$$10^{-3}$ & $7.62$ & $11.96$ & $3.74$ & $14.55$ & $2.42$ & $14.92$ & $-15.49$ & $-14.98$ & $-13.91$ \\ \cline{2-14}
 & g & $9.17\times10^{-1}$ & $9.19\times10^{-2}$ & $2.54$$\times$$10^{-1}$ & $8.11$ & $12.83$ & $2.90$ & $14.14$ & $2.87$ & $8.90$ & $-15.88$ & $-15.12$ & $-14.67$ \\ \cline{2-14}
 & r & $1.17$ & $7.30\times10^{-1}$ & $4.88$$\times$$10^{-3}$ & $9.09$ & $6.09$ & $3.23$ & $10.40$ & $1.64$ & $9.18$ & $-16.19$ & $-15.51$ & $-14.19$ \\ \cline{2-14}
 & i & $9.44\times10^{-1}$ & $2.16$ & $2.43$$\times$$10^{-2}$ & $6.66$ & $9.73$ & $3.39$ & $7.40$ & $1.96$ & $8.33$ & $-16.16$ & $-15.70$ & $-14.48$ \\ \cline{2-14}
 & z & $1.42\times10^{-1}$ & $2.09$ & $1.23\times10^{-1}$ & $5.15$ & $16.99$ & $3.21$ & $7.45$ & $2.61$ & $7.24$ & $-15.98$ & $-15.66$ & $-14.78$ \\ \cline{2-14}
 & y & $5.96\times10^{-1}$ & $2.21\times10^{-2}$ & $8.93$$\times$$10^{-2}$ & $8.43$ & $9.88$ & $2.09$ & $11.88$ & $2.02$ & $6.71$ & $-16.18$ & $-15.02$ & $-14.65$ \\  \hline
\multirow{6}{*}{SLy} & u & $2.66$$\times$$10^{-1}$ & $2.39$$\times$$10^{-1}$ & $4.10$$\times$$10^{-2}$ & $2.73$ & $10.90$ & $3.54$ & $12.30$ & $3.42$ & $15.05$ & $-14.81$ & $-14.40$ & $-14.14$ \\ \cline{2-14}
 & g & $3.38$$\times$$10^{-1}$ & $1.86$ & $1.19$$\times$$10^{-2}$ & $9.61$ & $10.52$ & $3.43$ & $7.18$ & $2.07$ & $11.45$ & $-15.97$ & $-15.44$ & $-14.17$ \\ \cline{2-14}
 & r & $1.46$ & $3.98$$\times$$10^{-3}$ & $1.04$$\times$$10^{-4}$ & $2.55$ & $5.91$ & $1.66$ & $9.85$ & $3.69$ & $10.61$ & $-15.49$ & $-14.11$ & $-11.86$ \\ \cline{2-14}
 & i & $9.36$$\times$$10^{-1}$ & $8.66$$\times$$10^{-2}$ & $3.20$$\times$$10^{-2}$ & $3.32$ & $6.30$ & $1.77$ & $7.84$ & $2.04$ & $8.64$ & $-15.84$ & $-14.83$ & $-14.45$ \\ \cline{2-14}
 & z & $1.84$ & $1.90\times10^{-1}$ & $2.58$$\times$$10^{-1}$ & $2.29$ & $4.92$ & $14.19$ & $10.66$ & $5.08$ & $8.63$ & $-15.47$ & $-15.44$ & $-14.99$ \\ \cline{2-14}
 & y & $3.41$$\times$$10^{-1}$ & $6.64\times10^{-1}$ & $2.00$$\times$$10^{-2}$ & $5.51$ & $7.43$ & $2.08$ & $6.59$ & $2.05$ & $9.15$ & $-16.10$ & $-15.18$ & $-14.35$ \\ \hline

\end{tabular}
\label{tab:t5}
\end{table*}

\subsection{Variation of nKLFs With Redshifts}
\label{subsec:LF_z}

The nKLFs at different redshifts can be slightly different because the distributions of BNS merger properties are different. As shown in Figure~\ref{fig:f10} for the BNS property distributions $P(m_{\rm tot}|z)$ and $P(q|m_{\rm tot},z)$ at different redshifts, there are slightly less BNS mergers with larger $m_{\rm tot}$ (e.g., $\gtrsim 3.0M_\odot$) at intermediate redshifts (e.g., $z=1$) than that at low redshift ($z=0.1$) under both the SLy and DD2 EOSs, but there are relatively more high-mass BNSs at higher redshifts $z=2$ than that at $z=1$. For high-mass BNS mergers $m_{\rm tot}>3.5M_\odot$, there are more asymmetric mergers at higher redshifts than those in the nearby universe, while for low-mass BNS ($m_{\rm tot}\lesssim 2.8M_\odot$) mergers, $P(q|m_{\rm tot},z)$ at different redshifts are more or less the same. However, the varying tendency of the BNS mass distribution dominates the variation of the nKLFs with redshifts. Therefore, the height of the brighter peak of the nKLF in each band at lower redshifts (e.g., $z=0.1$) is slightly lower than that at intermediate redshifts (e.g., $z=1$ or $z=1.5$), but the tendency is reversed at higher redshifts (e.g., $z=2$ or $z=3$) that the height of the brighter peak of the nKLF at intermediate redshifts (e.g., $z=1.5$) is relatively higher than that at higher redshifts (e.g., $z=2$ or $z=3$).

\section{Conclusions and Discussions}
\label{sec:conclusion}

In this paper, we investigate the dependence of kilonova luminosities on the parameters of BNS mergers and the distribution functions of kilonova peak luminosities (KLFs) at different bands as well as its dependence on the EOS of NSs by adopting a comprehensive semi-analytical model for kilonovae. In the kilonova model, we consider various engines that contribute to the kilonova emission, including the heating from the r process occurred in both the dynamical ejecta from the BNS merger and the wind ejecta from the remnant disc around the merger remnant, and the fallback accretion, etc. With this model, the observational u, g, i, r, y-, and z band LCs of GW170817 can be well matched and the observational upper limits on the magnitudes of the kilonova associated with GW190425  at different bands can be also consistently explained. We combine the ejecta properties predicted for BNS mergers under different EOSs by NR simulations, the BNS property distributions estimated from the population synthesis model for BNSs, and the best fit model for GW170817 (adopted to calibrate a number of model parameters for the BNS ejecta properties) to calculate the LCs for kilonovae resulting from general BNS mergers, estimate its peak luminosities at different bands, and thus obtain the KLFs. Our main conclusions are summarized as follows:

\begin{itemize}
\item We find that the heating of the free neutron skin can significantly boost the early-time luminosity of kilonova, and the fallback accretion heating of the dynamical ejecta can explain the  ``bump'' of the r and infrared band LCs. The multiband peak luminosities of AT2017gfo/GW170817 appear to be dominated by the r-process emission of the disc wind. While the contributions from the dynamical ejecta and the fallback accretion are expected to dominate the kilonova emission of GW190425, especially at the infrared bands and later time ($\gtrsim 0.5-1$\,day). 

\item The LCs of kilonova generated by a general BNS merger depend on the BNS properties. We find that (1) the mergers of BNSs with large total masses ($m_{\rm tot}$) generate kilonovae with lower peak luminosities and more rapid decay because of the much smaller masses of the resulting remnant discs compared with those of the mergers with small $m_{\rm tot}$; and (2) the mergers of BNSs with higher mass ratio (more asymmetric BNSs) generate kilonovae with higher peak luminosities as the more asymmetric BNS mergers tend to form remnant discs with larger masses, but this tendency is weaker for BNS mergers with low total mass since the masses of the remnant discs for these mergers appear not sensitive to the mass ratio. 

The LCs of kilonovae also depend on the EOS of NSs. Mergers of BNSs under a stiffer EOS generate kilonovae with higher peak luminosities than those under a softer EOS, simply because the former ones can eject more masses. More specifically, most kilonovae resulting from BNS mergers with $m_{\rm tot} \lesssim 3.1M_{\odot}$ under the DD2 EOS or $m_{\rm tot} \lesssim 2.8M_{\odot}$ under the SLy EOS have bright peak luminosities $M_{\rm peak}\lesssim -15.8/-15.5/-15.2$\,mag at the r, i, y, and z bands/g band/u band, mainly contributed by the wind ejecta from the large remnant discs. However, most kilonovae resulting from BNS mergers with $m_{\rm tot} \gtrsim 3.2M_{\rm \odot}$ under the DD2 EOS or $m_{\rm tot} \gtrsim 2.9M_{\rm \odot}$ under the SLy EOS have faint peak luminosities $M_{\rm peak}\gtrsim -15.2/-14.9/-14.5$\,mag at the r, i, y, and z bands/g band/u band because these mergers produce negligible remnant discs and its kilonova emission is mainly contributed by the dynamical ejecta typically with small masses. For the intermediate mass BNS mergers, their kilonovae peak luminosities are rather sensitive to $m_{\rm tot}$ and $q$. 

\item We obtain the normalized kilonova luminosity functions (nKLFs) with/out consideration of the selection effects on the viewing angles by GW detectors and use multiple Schechter functions to fit them. We find that the nKLFs are generally bimodal, with a bright component mainly contributed by BNS mergers with small $m_{\rm tot}$ ($\lesssim 3.2 M_\odot$/$2.8 M_\odot$ under the DD2/SLy EOS) and a faint component mainly contributed by BNS mergers with large $m_{\rm tot}$ ($\gtrsim 3.2 M_\odot$/$2.9M_\odot$ under the DD2/SLy EOS). The peak luminosity of a kilonova in the bright component is dominated by the emission from the wind ejecta, while that of a kilonova in the faint component is dominated by the emission from the dynamical ejecta. 

\item We find that the difference between the heights of the bright and faint components for the nKLF at each band obtained by assuming a stiff EOS (DD2) is substantially large than that obtained by assuming a soft EOS (SLy). The main reason is that BNS mergers with masses $\lesssim 3.2M_\odot$ mainly contribute to the bright components of the nKLFs under the DD2 EOS, while only those with masses $\lesssim 2.8M_\odot$ (substantially smaller than $3.2M_\odot$) contribute to the brighter component under the SLy EOS, which leads to relatively many more kilonovae in the bright components of nKLFs under the DD2 EOS than that under the SLy EOS.

\item Both the bright and faint components of the nKLFs, obtained by considering the GW selection effects on the viewing angles of kilonovae, are slightly narrower than those of the nKLFs obtained without considering such effects. The main reason is that kilonovae with large viewing angles and thus relatively faint peak luminosities may be missed by GW detectors. 
\end{itemize}

To obtain the above conclusions, we make a number of assumptions for the adopted kilonovae model, the BNS population synthesis model, the remnant disc and ejecta properties of BNS mergers, and the fallback accretion, which may introduce uncertainties to the estimates of kilonova LCs and the nKLFs as discussed below.

We mainly investigate the kilonova physics and estimate multiband KLFs and their dependence on the EOS in this paper. For this purpose, it is necessary to consider kilonovae resulting from BNS mergers distributed in a large parameter space. We adopt the semi-analytical model for kilonovae to make the analysis efficient and simpler over a large parameter space, which is based on the framework developed by many authors in the literature.
In the model, some ejecta properties given by NR simulations are adopted. However, many model parameters related to the kilonova emission components are assumed to be the same as the best-fitting ones to the LCs of GW170817, for simplicity. These parameters include the ejecting velocities of wind ejecta, opacities, floor temperatures, and geometry of different (blue, red, and purple) kilonova components, and they may be different for different kilonovae in reality. Therefore, the real nKLFs could be broader than those estimated in this work. 
In principle, one may adopt model grid for these parameters with more elaborate dynamical and radiative simulations rather than adopting the specific values constrained by GW170817 observations. It is worth noting that \citet{Pang202222058513} recently updated the multimessenger NMMA framework recently, with which the kilonova, GRB, afterglow, and GW signals can be analysed simultaneously. NMMA incorporates the kilonova model from the radiative transfer code POSSIS and was used to fit the AT2017gfo/GW170817 photometric data via the Gaussian Process Regression or neural networks, which is convenient and flexible for constraining the model parameters, and should be effective and efficient for the analysis of future observed kilonovae events. Based on the NMMA-like framework for kilonova, one may be able to construct vastly preferred model grid over a large parameter space for kilonova population analysis and predict various statistical properties of kilonovae, with calibrations by GW170817 and future multimessenger observations of other BNS mergers. Note that one may also directly adopt the constraints from observations (e.g., $R_{1.4M_\odot}$ by \citealt{Pang202222058513}) rather than a specified EOS, which should be more or less similar to the case adopting the SLy EOS in this paper. We defer this to future studies.

We adopt the simple formulas given by \citet{Perego-2017-37-37} and \citet{Arnett-1982-785-797} to estimate the intrinsic and emergent kilonova luminosities, instead of considering the detailed formation process of lanthanide and actinide elements, heating processes, and radiative transfer. It has been shown that the kilonova peak luminosities and the shape of kilonova LCs all depend on the detailed processes and may be highly diverse \citep[e.g.,][]{Zhu20219494, Korobkin2021116116, Barnes20214444, Wollaeger20211010}, which would also introduce some uncertainties to the estimates of kilonova properties and nKLFs. 

We note that in the above calculations the masses of the dynamical ejecta (equation~\ref{eq:mej_dyn}) and remnant disc (equation~\ref{eq:disc}) for different BNS mergers are inferred by combining the fitting formulas given by \citet{Coughlin-2019-91-96} and \citet{JuergenKrueger-2020-2002-7728} (Eqs.~\ref{eq:mej_dyn} and \ref{eq:disc}). 
Currently the fitting formulas given by different works are somewhat different 
and may have different systematic errors 
\citep[e.g.,][]{Dietrich-2017-105014-105014,Radice201836703682,Coughlin-2019-19-19, JuergenKrueger-2020-2002-7728}. Adopting the results from different simulations leads to different kilonova properties and the nKLFs. For BNS mergers with large total masses and large mass ratio, especially, the resulting dynamical ejecta inferred from different fitting formulas are either not well converged yet or even incorrect \citep{Henkel202222077658}, which may introduce substantial errors to the estimates of the nKLF faint components.
To clarify this, we further estimate the dependence of kilonova peak magnitudes on BNS properties and the KLFs by adopting either the COU formulas (Eqs.~\ref{eq:dyn_cou} and \ref{eq:disc_rea}) or the JKF formulas (Eqs.~\ref{eq:dyn_jkf} and \ref{eq:disc_cf}) as described below. Apparently, there are quite substantial differences in the predicted kilonova properties and luminosity functions by adopting different NR simulations results, which suggests that more accurate NR simulations are needed.

The COU formulas give intermediate dynamical ejecta mass ($\sim 0.001M_\odot$) and large disc mass ($\sim 0.2-0.3 M_\odot$) for BNS mergers at the low-mass end, but large dynamical ejecta mass (over $0.01M_\odot$ if the total mass  $M_{\rm tot}>3.5M_\odot$) and ignorable remnant disc mass (truncated at $M_{\rm tot}>2.7M_\odot$) for the high-mass BNSs. The dependence of the peak magnitudes on the mass ratio is weak. Therefore, the bright peaks of the KLFs mainly contributed by those more abundant low-mass BNSs via their disc winds, and the faint peaks are contributed by the less abundant high-mass BNSs via their dynamical ejecta. This leads to the distinctive two peaks in the luminosity functions, with the bright peak substantially higher than the faint peak.

When adopting the JKF formulas, the remnant disc mass depends not only on the BNS total mass but  also the mass ratio. If the BNS system is significantly asymmetric (with large $q$),  the remnant disc mass can still be large or intermediate even if the total mass as large as $M_{\rm tot}\sim 3.5M_\odot$, while the disc mass is substantially smaller at $M_{\rm tot}>3.0M_\odot$ if the system is close to symmetric (with $q\sim 1$). The dynamical ejecta mass estimated from the JKF formula decreases with increasing total mass and down to $<0.001M_\odot$ when $M_{\rm tot}>3.5M_\odot$, which contributes little to the kilonova luminosity. Therefore, there should be a significant number of kilonovae with intermediate peak luminosities, and kilonova produced by the high-mass BNSs dominates by the emission from the disc wind. The double peak structure of resulting KLFs should be less obvious in this case compared with those obtained by using the COU formulas.

We also note that the KLFs resulting from the JKF formulas cut at the bright end with magnitude $\sim-16.4$\,mag, slightly different from those obtained from the COU formulas ($\sim-16.0$\,mag). The reason is that the disc mass estimated from the COU formula is converged to $\sim0.2-0.3M_\odot$ with the BNS total mass decreasing to the low mass end, while that estimated from the JKF formula can be larger and do not strictly converge.

For the BNS population estimate, we adopt a simple recipe to relate the mass of the remnant NS with the CO-core mass, and we set the minimum mass of NSs as $1.15M_\odot$ and the maximum mass of NSs as the non-rotating TOV mass for each adopted EOS. The adopted linear relation between the NS mass and the CO-core mass may not accurately reflect the real one, which thus introduce some uncertainties to the estimates of the distributions of BNS properties. In addition, the setting of such a minimum NS mass is based on the current NS mass measurements, however, the theoretical allowed minimum NS mass could be smaller. By settings a different set of the minimum and maximum NS masses, the distributions of BNS merger properties obtained from the BNS population synthesis can be significantly different. We further check the effects of the minimum NS mass setting on the nKLFs by adopting $1M_\odot$ as the minimum one. We find that the fraction of kilonovae in the bright/faint component in this case is larger/smaller compared with the case of $1.15M_\odot$. The reason is that there are relatively more BNS mergers with smaller $m_{\rm tot}$ ($\lesssim 3.2M_\odot$/$2.8M_\odot$ for the DD2/SLy EOS) in the case with a smaller NS minimum mass. Considering rapidly rotating NSs, the maximum mass of NS could be somewhat larger than the TOV mass for any given EOS. In this case, the fraction of kilonovae in the faint/bright component would be larger/smaller compared with the case considered the TOV mass as the maximum NS mass.

To improve the estimates of kilonova properties and the KLFs for general BNS mergers, it is necessary to have more NR and radiative transfer simulations for a large number of BNS mergers with broad distributions of total mass and mass ratio under different EOSs. With these simulations, one may get more accurate fitting formulas for the properties of ejecta and discs of BNS mergers. One may also need to improve the BNS population synthesis model to give more accurate estimates for the property distributions of the BNS populations. Furthermore, future observations may detect many more GW events from the BNS mergers and find its associated kilonova phenomena, which can be used to constrain and calibrate the kilonova model parameters and their distributions, and thus help to deepen our understanding of not only the kilonova phenomenon but also the formation and evolution of the kilonova population. 

\section*{Acknowledgements}
We thank the referee for helpful comments and suggestions. This work is supported by the National Natural Science Foundation of China (grant nos. 11690024, 11873056, 11903030 and 12273035), the Strategic Priority Program of the Chinese Academy of Sciences (grant no. XDB 23040100), the National Key R\&D Program of China (grant nos. 2020YFC2201400 and 2021YFC2203100), and the Fundamental Research Funds for the Central Universities under grant nos. WK2030000036 and WK3440000004.


\section*{Data Availability}

The data underlying this article will be shared on reasonable request to the authors.

\bibliographystyle{mnras}
\bibliography{kilonovae_detecting}

\appendix
\section{The Posterior Distributions of the Model Parameters Fitted By the AT2017gfo Light Curves}

\begin{figure*}
\includegraphics[width=7in]{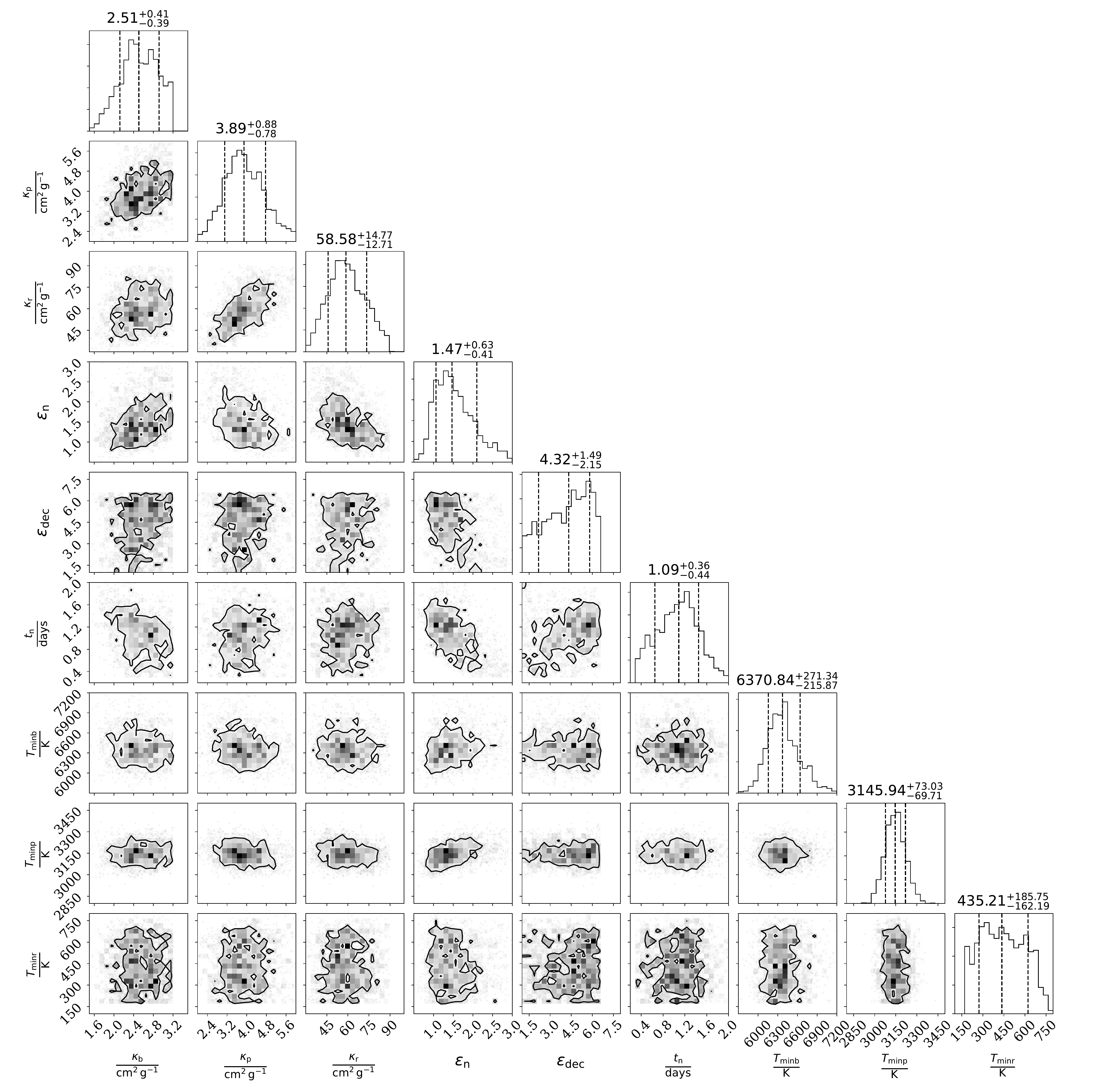}
\caption{
The corner plot for the posterior distributions of 9 of the $24$ model parameters (the opacities and the temperatures of the blue, purple and the red components, and the parameters of the free neutron skin of the dynamical ejecta) fitted with the LCs of AT2017gfo, and the best fits to the LCs are shown in Fig.~\ref{fig:f1} and the best fit parameters are listed in Table~\ref{tab:t1}. 
}
\label{fig:fa1}
\end{figure*}

\begin{figure*}
\includegraphics[width=7in]{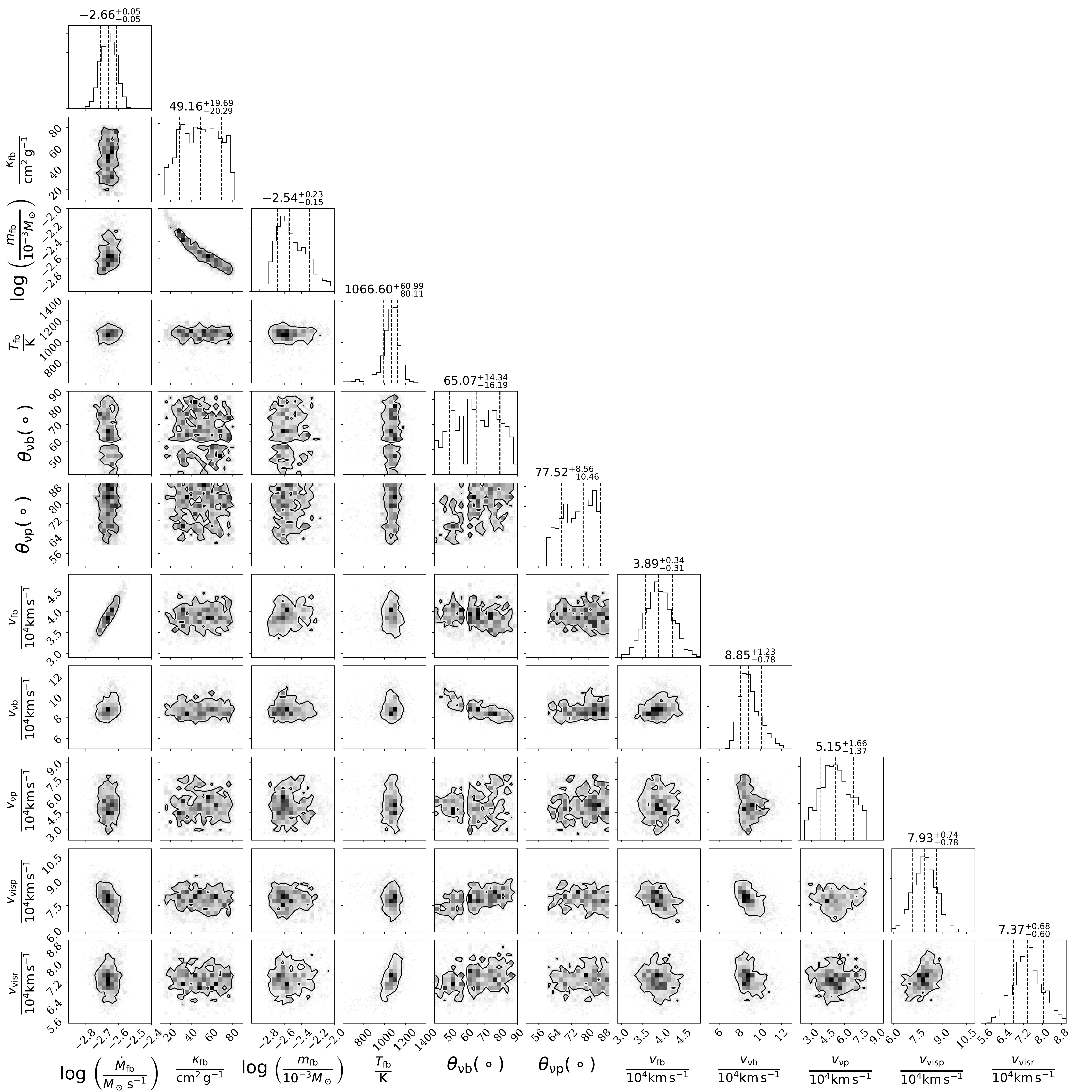}
\caption{
Similar to Fig.~\ref{fig:fa1}, but only showing the posterior distributions 11 of the $24$ model parameters (the accretion rate, the total mass, the opacity and the temperature of the fallback heating photosphere, and the related angles and the velocities parameters of disc wind).
}
\label{fig:fa2}
\end{figure*}

\begin{figure*}
\includegraphics[width=7in]{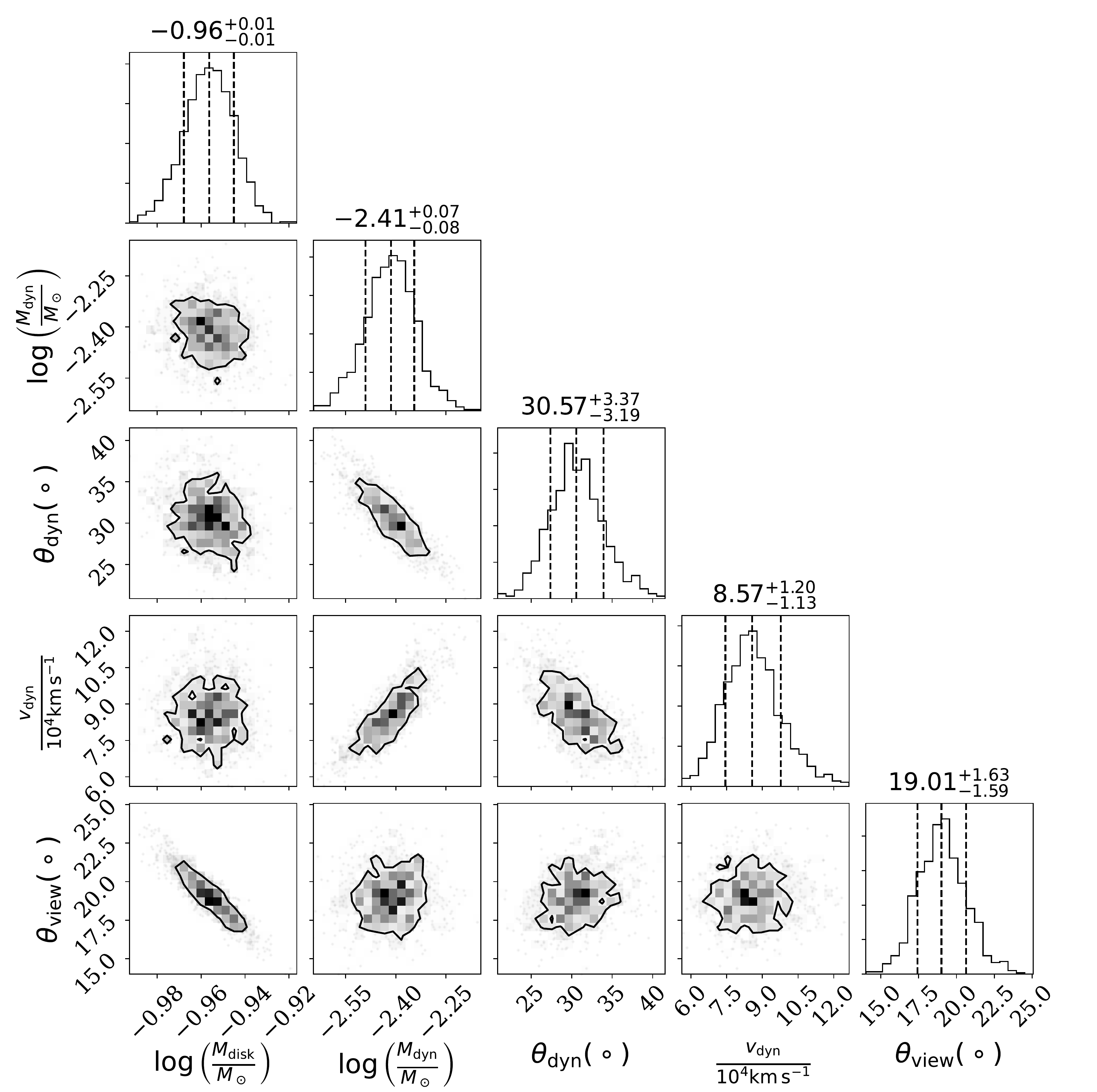}
\caption{
Similar to Fig.~\ref{fig:fa1}, but only showing the posterior distributions for $m_{\rm disc}$, $m^{\rm d}_{\rm ej}$, $v_{\rm dyn}$, $\theta_{\rm dyn}$, $\theta_{\rm view}$, respectively, for clarity. These five model parameters are set according to NR simulations for general BNS mergers.
}
\label{fig:fa3}
\end{figure*}

\bsp	
\label{lastpage}

\end{document}